\documentclass[aps,longbibliography, prl,showpacs,twocolumn,superscriptaddress,amsmath,amssymb,verbatim]{revtex4-2}
\usepackage[T1]{fontenc}
\usepackage[latin9]{inputenc}
\usepackage{bigints}
\usepackage[usenames,dvipsnames]{color}
\usepackage{float}
\usepackage{amssymb,amsmath}
\usepackage{graphicx}
\usepackage{epsfig}
\usepackage[normalem]{ulem}
\usepackage{booktabs}
\usepackage{setspace}
\usepackage{esint}
\usepackage{natbib}
\usepackage{bigints}
\usepackage{color}
\usepackage{babel}
\usepackage{amsmath,amsfonts,amssymb}
\usepackage{rotating}
\usepackage{graphicx}    
\usepackage{epstopdf}
\usepackage{color}
\usepackage[caption=false]{subfig}
\usepackage{soul}
\usepackage{color}
\usepackage{babel}
\usepackage{array}
\usepackage{float}
\usepackage{xfrac}
\usepackage{booktabs}
\usepackage{setspace}
\usepackage{esint}
\usepackage{chemformula}
\usepackage[unicode=true,pdfusetitle,
bookmarks=true,bookmarksnumbered=true,bookmarksopen=false,
breaklinks=true,pdfborder={0 0 0},pdfborderstyle={},backref=false,colorlinks=true]
{hyperref}
\hypersetup{
	linkcolor=red, citecolor=blue,  urlcolor=blue}
\definecolor{dark-blue}{rgb}{0.15,0.15,0.4}

\makeatletter
\@ifundefined{textcolor}{}
{%
 \definecolor{BLACK}{gray}{0}
 \definecolor{WHITE}{gray}{1}
 \definecolor{RED}{rgb}{1,0,0}
 \definecolor{GREEN}{rgb}{0,1,0}
 \definecolor{BLUE}{rgb}{0,0,1}
 \definecolor{CYAN}{cmyk}{1,0,0,0}
 \definecolor{MAGENTA}{cmyk}{0,1,0,0}
 \definecolor{YELLOW}{cmyk}{0,0,1,0}
}

\newcommand{\SPHIDE}[1]{{}}

\makeatother

\usepackage{babel}
\begin{document}

	\title{A novel Gapless Quantum Spin Liquid in the $S = 1$ $4d^{4}$-honeycomb material Cu$_{3}$LiRu$_{2}$O$_{6}$ }
	
	\author{Sanjay Bachhar}
	\thanks{\href{mailto:sanjayphysics95@gmail.com}{sanjayphysics95@gmail.com}}
	\affiliation{Department of Physics, Indian Institute of Technology Bombay, Powai, Mumbai 400076, India}
	
	\author{Nashra Pistawala} 
	\affiliation{Department of Physics, Indian Institute of Science Education and Research, Pune, Maharashtra-411008, India}
	
		\author{S. Kundu}
	\affiliation{Department of Physics, Indian Institute of Technology Madras, Chennai 600036, India}

	\author{Maneesha Barik}
	\affiliation{Department of Physics, Indian Institute of Technology Madras, Chennai 600036, India}

	\author{M. Baenitz}
	\affiliation{Max Planck Institute for Chemical Physics of Solids, 01187 Dresden, Germany}
	
	\author{Jörg Sichelschmidt }
	\affiliation{Max Planck Institute for Chemical Physics of Solids, 01187 Dresden, Germany}

	\author{Koji Yokoyama }
	\affiliation{ISIS Pulsed Neutron and Muon Source, STFC Rutherford Appleton Laboratory,Harwell Campus, Didcot, Oxfordshire OX110QX, United Kingdom}
		\author{P. Khuntia}
	\affiliation{Department of Physics, Indian Institute of Technology Madras, Chennai 600036, India}
	\affiliation{Quantum Centre of Excellence for Diamond and Emergent Materials,
		Indian Institute of Technology Madras, Chennai 600036, India.}

	\author{Surjeet Singh} 
	\affiliation{Department of Physics, Indian Institute of Science Education and Research, Pune, Maharashtra-411008, India}

	\author{A.V. Mahajan}
	 \thanks{\href{mailto:mahajan@phy.iitb.ac.in}{mahajan@phy.iitb.ac.in}}
	\affiliation{Department of Physics, Indian Institute of Technology Bombay, Powai, Mumbai 400076, India}

\date{\today}
\begin{abstract}

 We report discovery of a novel gapless quantum spin liquid in the $S = 1$ honeycomb system \ch{Cu3LiRu2O6} with Ru$^{4+}$ ($4d^{4}$) where moments remain dynamic down to 50 mK. Heat capacity measurements show no sign of magnetic ordering down to 60 mK inspite of a Curie-Weiss temperature, $\theta_{CW}=-222$ K -indicating a strong antiferromagnetic interaction. In zero field, magnetic heat capacity, $C_{m}$ shows a linear $T$-dependence with Sommerfeld coefficient $\gamma= 107 $ mJ/mol K$^{2}$ which is much larger than that found in typical Fermi liquids. Our local probe $^7$Li nuclear magnetic resonance (NMR) measurements find a significant temperature-independent $^{7}$Li NMR shift (and hence a non-zero spin susceptibility) at low-$T$ and a linear $T$-variation of the $^{7}$Li NMR spin-lattice relaxation rate 1/$T_{1}$ at low-$T$ reminiscent of fermionic excitations. Muon spin relaxation ($\mu$SR) measurements detect neither long range ordering nor spin freezing down to 50 mK and the temperature variation of the muon depolarization rate $\lambda$ shows a gradual increase with decreasing temperature and a leveling off below about 1 K evincing a persistent spin dynamics  common to several spin liquid candidates. Our results provide strong signatures of a quantum spin liquid in the titled honeycomb material.

\end{abstract}

\maketitle

\textit{Introduction } Excitonic novel magnetism in $4d^{4}$ based systems is of great interest and subject of intense investigation worldwide \cite{Khaliullin2013_prl, Khaliullin2019_prb, Khaliullin2019_prl, Khaliullin_2010, Takagi2022}. In the Kitaev honeycomb lattice of \ch{A2IrO3} (A=Na, Li) \cite{Coldea2016,Takagi2019,Yogesh2010,Freund2016,Choi2012} spin-orbit coupling (SOC) is dominant and on-site electronic correlations are significantly reduced because of the extended nature of the $5d$ orbitals. A balance between the competing energies might be expected in the $4d$-based systems with the possibility of exotic states rarely seen in other materials \cite{Ghosh2021,AVM2016,RKumar2019,Takagi2022}. Excitonic magnetism as predicted  by Khaliullin \cite{Khaliullin2013_prl, Khaliullin2019_prl} can be experimentally realized if there is sufficient SOC to stabilize a total angular momentum $J_{\text{eff}}=1$ state. Accordingly, honeycomb structure decorated with any of the $d^{4}$ ions (Ru$^{4+}$, Re$^{3+}$, Os$^{4+}$, and Ir$^{5+}$) has been proposed to manifest novel physical properties \cite{Khaliullin2013_prl}.   Absence of frustration in a spin angular momentum $S=1$ honeycomb system without bond anisotropy, often leads to magnetic ordering \cite{RKumar2019ALMO}. As mentioned in \cite{Lee2013} despite its bipartite nature, low  coordination number (z=3) of the honeycomb lattice increases the quantum fluctuations and may give rise to a spin liquid as a competing ground state. Also, there may be a few  ways to introduce frustration in $4d^4  (S=1)$ honeycomb systems. Orbital frustration can be a source of frustration in the spin channel and might open up the possibility of realizing a spin-orbital liquid with both spin and orbital entanglement \cite{Trivedi2017}. The possible importance of further neighbor interactions as also bi-quadratic and ring-exchange terms can as well give rise to frustration in the system \cite{RKumar2019}. However, quantum spin liquid behavior has not been conclusively observed experimentally in $4d^4$ honeycomb systems so far. One of the promising honeycomb systems containing Ru as the $4d^4$ ($S=1$) magnetic atom is \ch{Ag3LiRu2O6} \cite{RKumar2019} where the spins remains on the borderline between static and dynamic even at 20 mK. Takagi \textit{et al.} \cite{Takagi2022} reported that \ch{Ag3LiRu2O6} with $4d^4$ Ru$^{4+}$ ions at ambient pressure forms a honeycomb lattice of spin-orbit-entangled singlets,
which is a playground for frustrated excitonic magnetism. Replacing inter-layer atom does affect magnetism and associated low-energy excitations in frustrated 2D layered systems \cite{Kitagawa2018, Singer2019}. With the above background in mind, we set out to investigate \ch{Cu3LiRu2O6} (CLRO) where the lighter Cu (which is in the nonmagnetic Cu$^{1+}$ state) replaces the relatively heavier Ag at the inter-layer sites which we found results in a novel gapless quantum spin liquid.

\begin{figure}[h]
	\centering{}\includegraphics[width=1.0\linewidth]{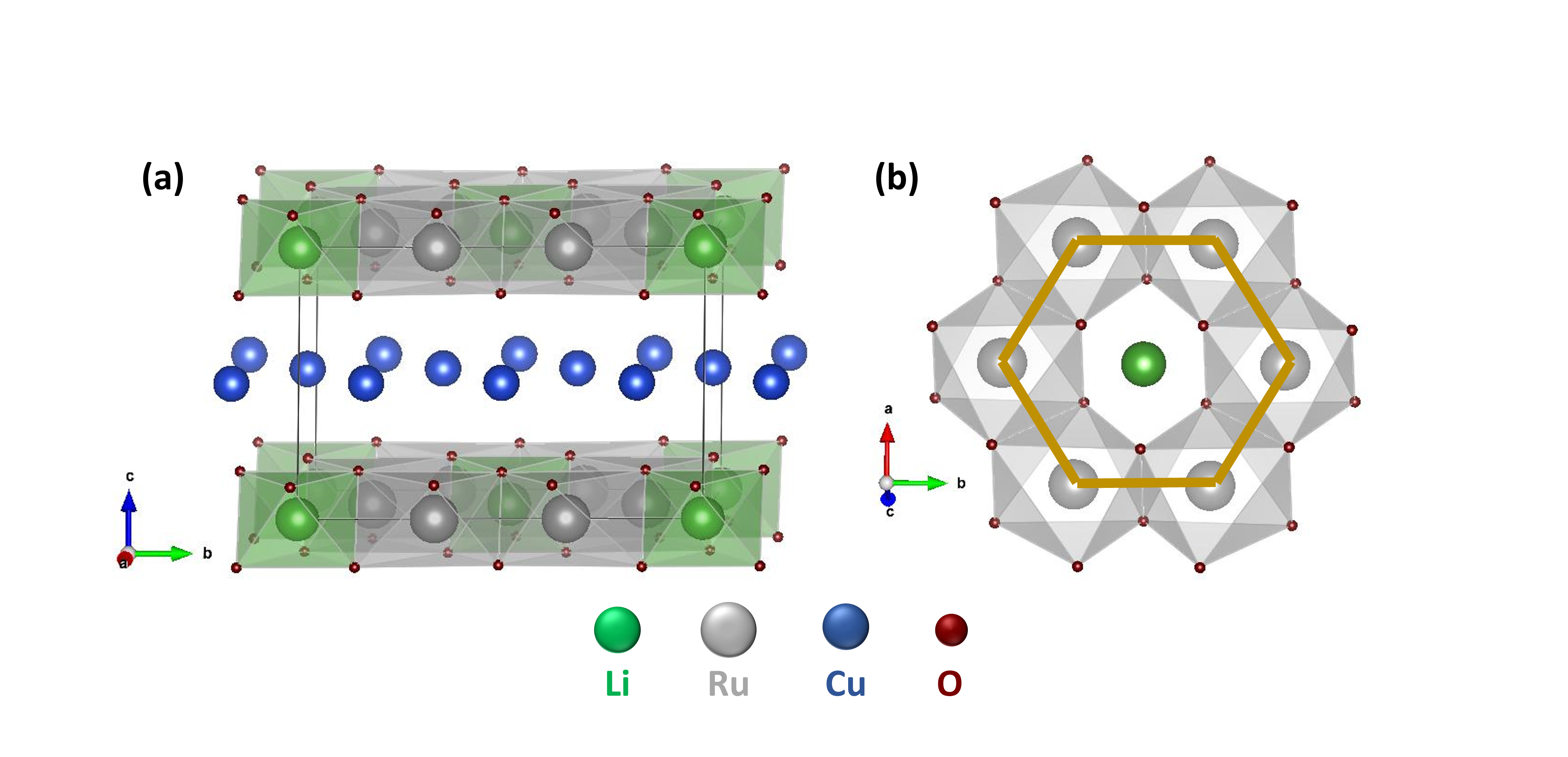}\caption{\label{fig:CLRO-schematic}{\small{}Schematic crystal structure of Cu$_{3}$LiRu$_{2}$O$_{6}$. (a) 3D view (b) Ru-honeycomb plane}}
	\label{CLRO_crystal}
\end{figure}
 
Before going into the details of magnetism of CLRO, we discuss a few important structural aspects. CLRO crystallizes in the $C2/m$ space group similar to its sister compound \ch{Ag3LiRu2O6}. A schematic crystal structure is shown in Figure \ref{CLRO_crystal}. A lighter Cu atom between the magnetic planes compared to Ag reduces the distance between the two honeycomb layers to 6.05(2) \AA  \, (Cu-compound) from 6.51(0)\AA  \, (Ag-compound) and is expected to influence the underlying magnetism. Structural analysis suggests a nearly perfect honeycomb with a nearest-neighbor Ru-Ru distance of 2.98 {\AA} and a Ru-O-Ru bond angle of 94.9{\textdegree} (for details see SM \cite{SM}). Our X-band electron spin resonance (ESR) measurements could not reveal ESR active probe spins, which suggests that Cu is non-magnetic (Cu$^{1+}$) in the system and the magnetism arises from Ru$^{4+}$ only.


Our bulk magnetization, heat capacity, nuclear magnetic resonance (NMR) and muon spin relaxation ($\mu$SR) measurements show no signature of long-range magnetic ordering down to 50 mK inspite of a Curie-Weiss temperature of approximately $-$222 K -indicating  strong antiferromagnetic interactions between Ru$^{4+}$ moments.  Temperature independent variation of $^{7}$Li NMR shift ($^{7}K$) below 200 K along with a power law ($\sim T$) in $^{7}$Li NMR spin-lattice relaxation rate, 1/$T_{1}$ at low-$T$ indicates Korringa-like behavior $1/T_{1}T \sim K^{2}$ as expected for a Fermi liquid. This is also consistent with the $T$-linear variation of the zero-field specific heat. A large Sommerfeld coefficient $\gamma$ = 107 mJ/mol K$^2$ is inferred in zero field. In applied magnetic fields, the specific heat variation morphs into a power law variation with temperature ($\sim$ $T^{1.8}$) indicating gapless excitations. Data collapse/scaling of the specific heat with $T/B$ is observed which originates from possible formation of random singlets due to a small amount of defects whereas the bulk phase is a QSL. The muon spin relaxation rate, $\lambda$ becomes nearly $T$-independent below about 1 K (down to 50 mK), indicating the presence of persistent dynamics in the system. 

\begin{figure}[h]
	\centering{}\includegraphics[scale=0.33]{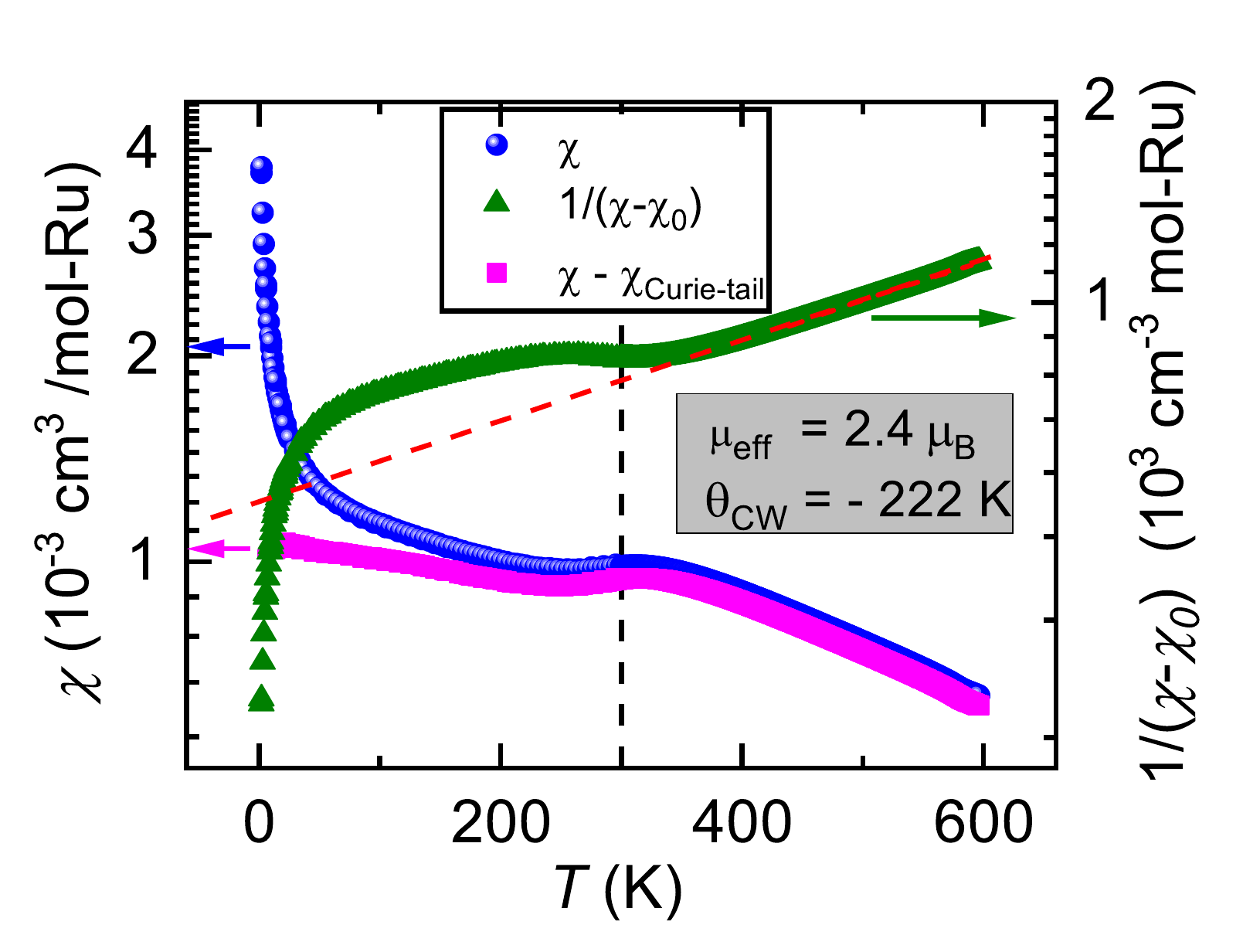}\caption{\label{fig:Curie-Weiss-fit-of CLRO}{\small{} The left $y$-axis shows the temperature dependence of the susceptibility $\chi(T) = \frac{M(T)}{H}$  of \ch{Cu3LiRu2O6}.  Also shown is the susceptibility after subtraction of a Curie-tail $\chi(T)-\chi_{Curie-tail}$. The right $y$-axis shows the inverse susceptibility  free from a  temperature independent contribution $\chi_{0}$ and the dashed line is a linear fit.}}
\end{figure}

\textit{Results and discussion } We observe that the bulk magnetic susceptibility (Figure \ref{fig:Curie-Weiss-fit-of CLRO}) increases from 600 K to 300 K with a Curie-Weiss variation. There are no anomalies down to 2 K (except for the 300 K broad anomaly, consistent with $^{7}$Li NMR shift $^{7}K$ shown later but not associated with long range ordering), which rules out the presence of any long-range or short-range ordering within the system.
 Fitting the high-temperature part (400-600 K) with the Curie-Weiss law ($\chi = \chi_{0} + \frac{C}{T-\theta_{CW}}$, where $\chi_{0}$, \textit{C}, and $\theta_{CW}$ are the temperature-independent susceptibility, Curie constant, and Curie-Weiss temperature representing a characteristic energy scale, respectively) yields $\chi_{0} = 1.0 × 10^{-6}$ (cm$^{3}$/mol Ru), \textit{C} = 0.71 (K cm$^{3}$/mol Ru), and $\theta_{CW} = -222$ K -indicating  strong antiferromagnetic interactions. The effective moment was found to be $\mu_{\text{eff}} = \sqrt{8C}\,\mu_{B}= 2.4 \mu_{B}$,  which is close to that expected for $g=2$ and $S=1$ ($\mu_{\text{eff}} = 2.83  \mu_{B}$). The susceptibility after subtraction of a Curie-tail \footnote{Low-$T$ Curie-tail is from about 3\% S=1/2 defects impurity.} ($\chi(T)-\chi_{Curie-tail}$) is shown in Figure \ref{fig:Curie-Weiss-fit-of CLRO}, and is nearly temperature independent which is consistent with temperature independent $^{7}$Li NMR-shift $^{7}K$ (shown later). The inverse of the susceptibility (free from the temperature-independent part) is shown on the right \textit{y}-axis of Figure \ref{fig:Curie-Weiss-fit-of CLRO}. The frustration parameter ($f = \frac{\mid\theta_{CW}\mid}{T_{N}} \ge 4440$) (assuming $T_{N}$ to be below the lowest temperature of 50 mK pertaining to $\mu$SR measurements) is very high, indicating that CLRO is a highly frustrated system.  
\begin{figure}[h]
	\centering{}\includegraphics[scale=0.33]{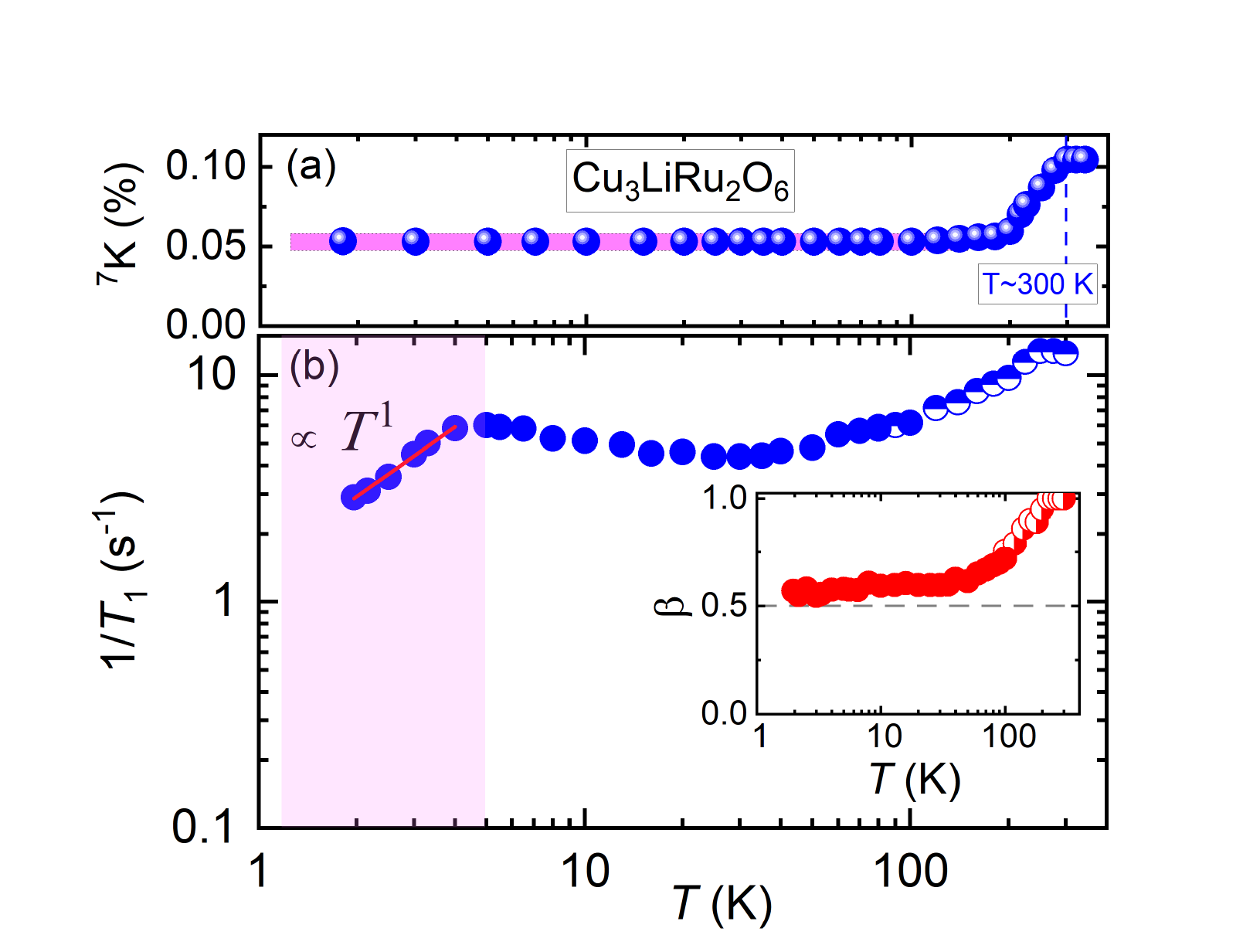}\caption{\label{fig:KT1_CLRO}{\small{} (a) $^7$Li NMR Knight Shift, $^{7}K$ as a function of temperature (2-340 K) for \ch{Cu3LiRu2O6}. The temperature-independent, Pauli-like $^{7}K$ below about 200 K is reminiscent of a quantum spin liquid.  (b) $^{7}$Li spin lattice relaxation rate 1/$T_{1}$ as a function of temperature for \ch{Cu3LiRu2O6}.   The variation of the stretching exponent $\beta$ with $T$ is shown in the inset.  }}
\end{figure}

Next, we present results of NMR which is a powerful local probe of intrinsic spin susceptibility (in the present case, through the $^{7}$Li NMR shift, $^{7}K$) and low-energy excitations (through the $^{7}$Li NMR spin-lattice relaxation rate, 1/$T_{1}$). Figure \ref{fig:KT1_CLRO}(a) shows $^{7}K$ as a function of temperature for a randomly oriented powder sample of CLRO. We find that $^{7}K$ decreases from its room temperature value to a constant below about 200 K. This implies that the intrinsic spin susceptibility is Pauli-like at low-$T$ which is reminiscent of a quantum spin liquid. This is qualitatively similar to the variation in \ch{H3LiIr2O6} (HLIO) \cite{Kitagawa2018}. 
 Figure \ref{fig:KT1_CLRO}(b) shows the $^{7}$Li NMR spin lattice relaxation rate, 1/$T_{1}$ as a function of temperature.  1/$T_{1}$ measures the $q$-averaged imaginary part of the dynamical spin susceptibility, $\chi^{\prime\prime}(q,\omega_{0})$ \footnote{Dynamical spin susceptibility is a measure of the response of a material's magnetic moment to an applied magnetic field. Specifically, it is the Fourier transform of the auto-correlation function of the spin fluctuations in the material.}. We fitted the recovery of the longitudinal nuclear magnetization \textit{M$_{z}$(t$_{d}$)} -measured by a saturation recovery method- with a stretched exponential function to extract 1/$T_{1}$ and the stretching exponent $\beta$. The variation of the stretching exponent $\beta$ with $T $ is shown in Figure \ref{fig:KT1_CLRO}(b) (inset).   Above 250 K, $\beta$ is 1, however, below 250 K, $\beta$ decreases and becomes nearly constant ($\sim$ 0.6) below about 60 K. This deviation of $\beta$ from the ideal value ($=1$) is possibly caused by the impact of magnetic defects that are distributed randomly in the host spin-lattice. In frustrated magnets such as quantum spin liquids, a value of $\beta \sim 0.5-0.6$ has been commonly observed \cite{Kitagawa2018}, which corresponds to randomized dipolar hyperfine fields. The linear power law variation ($\sim T^{1}$) of $^{7}$ Li NMR spin-lattice relaxation rate 1/$T_{1}$ at low-$T$ indicates gapless excitations, often seen in  quantum spin liquids\cite{Fermi-Liquid-1,Fermi1996}. 
  \begin{figure}[h]
  	\centering{}\includegraphics[scale=0.33]{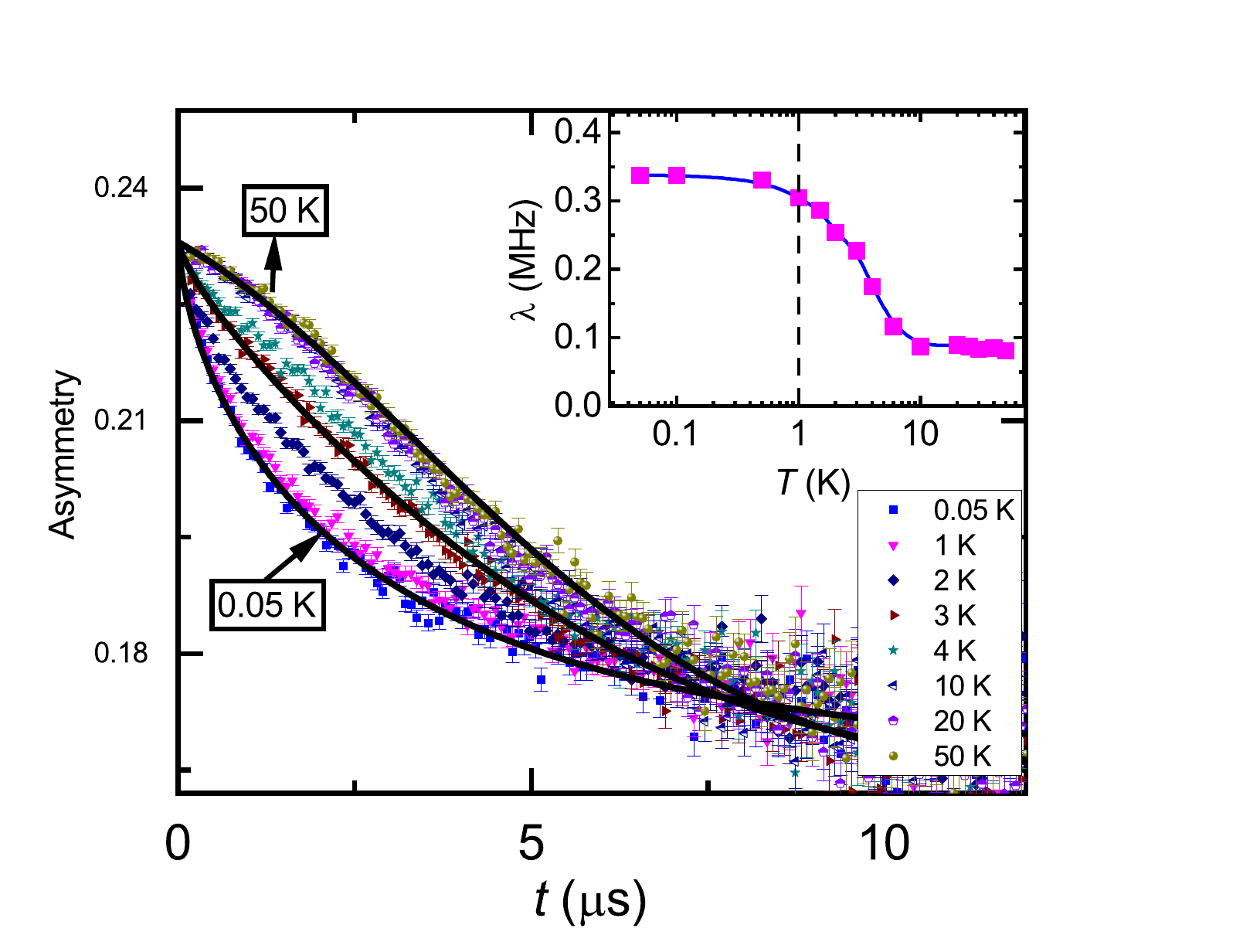}\caption{\label{fig:ZF_asym_CLRO}{\small{} The variation of the zero-field muon asymmetry with time is shown in the temperature range 50 mK and down to 50 K for \ch{Cu3LiRu2O6}. (Inset) On fitting these data as described in the text, the muon depolarisation rate $\lambda$ was obtained and variation with temperature is shown. The vertical dashed line ($T = 1$ K) is a guide to the eye, below which $\lambda$ becomes constant inferring persistent spin dynamics.}}
  \end{figure} 
  
  Another piece of evidence we now present is that of local moment dynamics in CLRO through $\mu$SR measurement which were carried out at the ISIS Facility, UK. Zero-field $\mu$SR data show no oscillations down to 50 mK which is  evidence of the absence of long-range magnetic order consistent with our other experiments. We could reliably fit the muon asymmetry using $ A(t)= A_{rel} G_{KT}(\Delta,T) exp(- \lambda t) + A_{0} $. Here, $G_{KT}(\Delta,T)$ is the Kubo-Toyabe function which models the relaxation of muons in a Gaussian distribution of magnetic fields from nuclear moments and $A_{rel}$ is the relaxing asymmetry. From these fits at high-$T$ (where the nuclear moments dominate the relaxation), we obtain $\Delta$ which corresponds to a field distribution of about 1.8 Oe. This value is typical of nuclear dipolar fields at the muon site, in the present case arising from $^{63,65}$Cu, $^{6,7}$Li, and $^{99,101}$Ru nuclei. The exponential term $exp(- \lambda t)$ arises from the relaxation due to fluctuations of the electronic local moments. Below 3 K, the data were well fitted with a stretched exponential in addition to a constant i.e., $A(t)=A_{rel} exp(-(\lambda t)^\beta) + A_{0}$. Here, $\beta$ is the stretching exponent which is often seen in non-uniform magnetism. Below about 3 K, $\beta$ deviates from 1 and becomes constant with value $\sim$ 0.7 below 1 K and down to 0.05 K together with the muon spin relaxation rate $\lambda$ which remains constant as well. This indicates the persistence of local moment dynamics in the system. We have also performed longitudinal field (LF) measurements at various fields and find that at our highest field (4500 G), a residual relaxation is still present (see SM \cite{SM}).

  \begin{figure}[h]
  	\centering{}\includegraphics[scale=0.33]{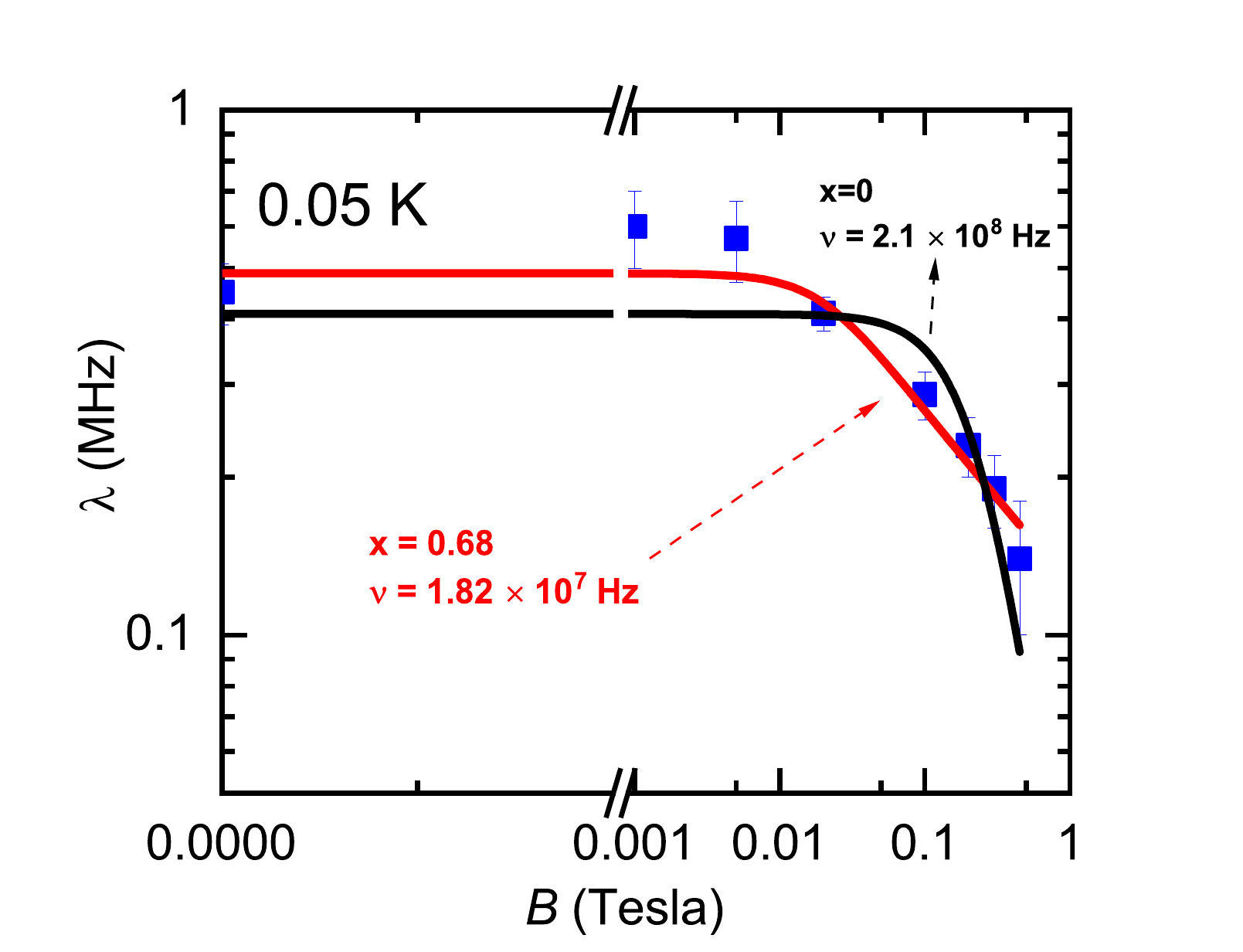}\caption{\label{fig:lamda_H_CLRO}{\small{}The variation of the muon relaxation rate at 50 mK is shown as a function of the longitudinal magnetic field for \ch{Cu3LiRu2O6}. The red curve is a fit to Equation \ref{lamda_H} with $x$ as a fitting parameter and the black curve is a fit to the same equation but with $x = 0$. }}
  	\label{lamda_H_CLRO}
  \end{figure} 
 Following the analysis of the field dependence of $\lambda$ as in Ref. \cite{YMGO}, we fit the
data to the following equation:

\begin{equation}
	\lambda (H) = 2 \Delta ^2 \tau ^x \int_{0}^{\infty} t^{-x} exp(- \nu t) Cos (\gamma_{\mu} H t) dt
	\label{lamda_H}
\end{equation}

where $\nu$ is the fluctuation frequency of local moments and
$\Delta$ is the distribution width of the local magnetic fields.
The muon gyromagnetic ratio is 
$\gamma_{\mu}$ = 2$\pi$ x 135.5342
MHz/T. A fit with $x = 0$ (black curve in Figure \ref{lamda_H_CLRO})
which implies an exponential auto-correlation function
$S(t) \sim exp(-\nu t)$ does not fit the data well and rather
$S(t) \sim (\tau/t)^x exp(- \nu t)$ is needed to fit the data. The
red solid curve is a fit to Equation \ref{lamda_H} and gives $x = 0.68$ and $\nu = 1.82 \times 10^7$ Hz. $\tau$ is the early time cut-off and is fixed
to $10^{-12} s$. The LF dependence shows a striking resemblance to that observed in \ch{YbMgGaO4} \cite{YMGO} and \ch{Sr3CuSb2O9} \cite{SKundu2020}, characterized by a local moment fluctuation frequency of approximately 18 MHz and the existence of spin correlations over long-time. The qualitative and quantitative results obtained through $\mu$SR analysis of CLRO align with those observed in other quantum spin liquid  candidates \cite{SKundu2020}.

\begin{figure}[h]
	\centering{}\includegraphics[scale=0.33]{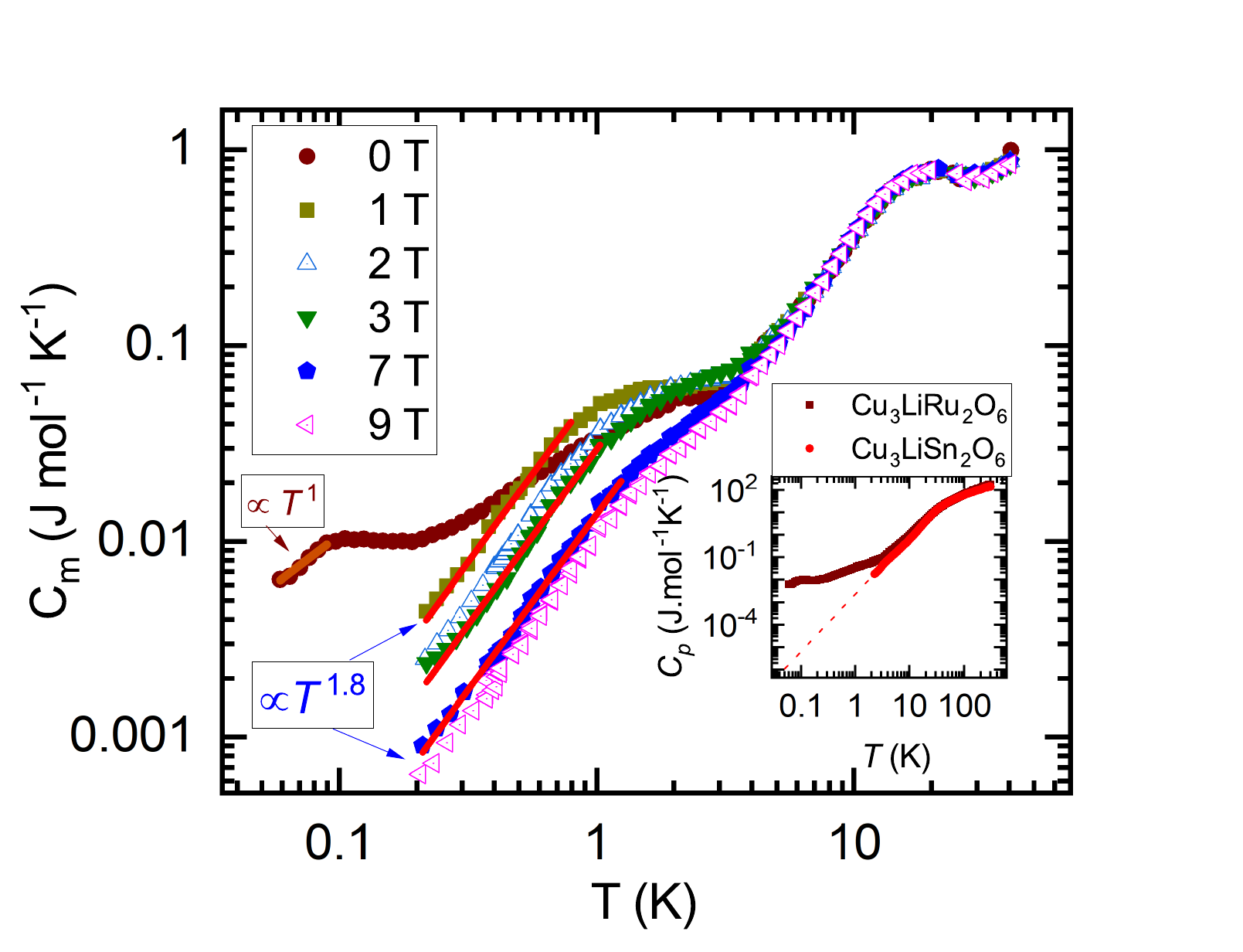}\caption{\label{fig:Cp_CLRO}{\small{}Magnetic heat capacity, $C_{m}$ with $T$ at various fields 0-9 Tesla. In zero field, linear ($\sim$ $T$) variation is observed. With application of field, nearly quadratic ($\sim$ $T^{1.8}$) power law variation is observed.(Inset) Specific heat, C$_{p}$ as a function of temperature for \ch{Cu3LiRu2O6} (brown circle) and its non-magnetic analog Cu$_{3}$LiSn$_{2}$O$_{6}$ (red line).}}
\end{figure}

Finally, we discuss the heat capacity \textit{C$_{p}$(T)} of CLRO as presented in Figure \ref{fig:Cp_CLRO}. The magnetic heat capacity \textit{C$_{m}$(T)} was obtained by subtracting the \textit{C$_{p}$(T)} of a  non-magnetic analog, Cu$_{3}$LiSn$_{2}$O$_{6}$  (for details see SM \cite{SM}). In zero magnetic field, no sharp anomaly is seen in \textit{C$_{m}$(T)} down to 60 mK (Figure \ref{fig:Cp_CLRO}) consistent with the absence of long-range magnetic order as was concluded from our muSR experiments. Our NMR experiments  further corroborate this conclusion. In zero field, $C_{m} = \gamma T$ with Sommerfeld coefficient, $ \gamma=107$ mJ/mol K$^{2}$ at low-$T$ (the corresponding value in a metal would be two orders of magnitude smaller). This indicates gapless fermion-like excitations which is consistent with the $T$-linear NMR spin lattice relaxation rate.

 Power law variation of $C_{m}$ at low-$T$ indicates gapless excitations which is consistent with other experiments. On application of a field, the exponent changes from 1 to 1.8 which might be due to vacancy-induced states in the quantum spin liquid. At low-$T$, $C_{m}$ is dependent on the applied field and scaling of $C_{m}$ data (See SM\cite{SM}) is observed which could be due to the effect of the field on the defect-induced low-energy density-of-states. We speculate that the scaling originates from a tiny amount of vacancy presence in the titled material and leads to possible local random singlet-like phase on top of a quantum spin liquid phase.  

\textit{Conclusion.- } We propose CLRO to be an $S=1$ (Ru$^{4+}$-based) gapless spin liquid which shows an absence of long-range ordering down to 50 mK. We observe a Pauli-like, $T$-independent NMR shift (non-zero spin susceptibility) which is reminiscent of a quantum spin liquid. The $T$-linear variation in zero field-$C_{m}$  and NMR-1/$T_{1}$ indicates gapless excitations. The data collapse/scaling in $C_{m}$ is suggestive of random singlet formation on top of the QSL state. Our results provide strong evidence of a QSL state with fermionic-like spin excitations in a promising $4d^{4}$ honeycomb system which will further motivate experimental and theoretical work in establishing a realistic Hamiltonian and will provide insights into unconventional low-energy excitations in the spin liquid state of this class of 4d honeycomb materials. This also provides motivation to carry out inelastic neutron scattering measurements for further validating our conclusions.      
\section{acknowledgment}

We thank MOE India, STARS project ID:358 for financial support.  We thank Central Facilities at IIT Bombay for support for various measurements.
Experiments at the ISIS Neutron and Muon Source
were supported by a beam-time allocation RB2310046 from the Science
and Technology Facilities Council (https://doi.org/10.5286/ISIS.E.RB2310046-1). P.K. acknowledges the funding by the Science and
Engineering Research Board, and Department of Science and Technology, India through Research Grants. SK thanks ANRF, Govt. of India through NPDF research grant PDF/2022/003836. We thank Indra Dasgupta, Sumiran Pujari, and Ritwik Das for useful discussions.
\bibliographystyle{apsrev4-2}
\bibliography{citation}

\begin{thebibliography}{29}%
\makeatletter
\providecommand \@ifxundefined [1]{%
 \@ifx{#1\undefined}
}%
\providecommand \@ifnum [1]{%
 \ifnum #1\expandafter \@firstoftwo
 \else \expandafter \@secondoftwo
 \fi
}%
\providecommand \@ifx [1]{%
 \ifx #1\expandafter \@firstoftwo
 \else \expandafter \@secondoftwo
 \fi
}%
\providecommand \natexlab [1]{#1}%
\providecommand \enquote  [1]{``#1''}%
\providecommand \bibnamefont  [1]{#1}%
\providecommand \bibfnamefont [1]{#1}%
\providecommand \citenamefont [1]{#1}%
\providecommand \href@noop [0]{\@secondoftwo}%
\providecommand \href [0]{\begingroup \@sanitize@url \@href}%
\providecommand \@href[1]{\@@startlink{#1}\@@href}%
\providecommand \@@href[1]{\endgroup#1\@@endlink}%
\providecommand \@sanitize@url [0]{\catcode `\\12\catcode `\$12\catcode
  `\&12\catcode `\#12\catcode `\^12\catcode `\_12\catcode `\%12\relax}%
\providecommand \@@startlink[1]{}%
\providecommand \@@endlink[0]{}%
\providecommand \url  [0]{\begingroup\@sanitize@url \@url }%
\providecommand \@url [1]{\endgroup\@href {#1}{\urlprefix }}%
\providecommand \urlprefix  [0]{URL }%
\providecommand \Eprint [0]{\href }%
\providecommand \doibase [0]{https://doi.org/}%
\providecommand \selectlanguage [0]{\@gobble}%
\providecommand \bibinfo  [0]{\@secondoftwo}%
\providecommand \bibfield  [0]{\@secondoftwo}%
\providecommand \translation [1]{[#1]}%
\providecommand \BibitemOpen [0]{}%
\providecommand \bibitemStop [0]{}%
\providecommand \bibitemNoStop [0]{.\EOS\space}%
\providecommand \EOS [0]{\spacefactor3000\relax}%
\providecommand \BibitemShut  [1]{\csname bibitem#1\endcsname}%
\let\auto@bib@innerbib\@empty
\bibitem [{\citenamefont {Khaliullin}(2013)}]{Khaliullin2013_prl}%
  \BibitemOpen
  \bibfield  {author} {\bibinfo {author} {\bibfnamefont {G.}~\bibnamefont
  {Khaliullin}},\ }\href {https://doi.org/10.1103/PhysRevLett.111.197201}
  {\bibfield  {journal} {\bibinfo  {journal} {Phys. Rev. Lett.}\ }\textbf
  {\bibinfo {volume} {111}},\ \bibinfo {pages} {197201} (\bibinfo {year}
  {2013})}\BibitemShut {NoStop}%
\bibitem [{\citenamefont {Chaloupka}\ and\ \citenamefont
  {Khaliullin}(2019)}]{Khaliullin2019_prb}%
  \BibitemOpen
  \bibfield  {author} {\bibinfo {author} {\bibfnamefont {J.~c.~v.}\
  \bibnamefont {Chaloupka}}\ and\ \bibinfo {author} {\bibfnamefont
  {G.}~\bibnamefont {Khaliullin}},\ }\href
  {https://doi.org/10.1103/PhysRevB.100.224413} {\bibfield  {journal} {\bibinfo
   {journal} {Phys. Rev. B}\ }\textbf {\bibinfo {volume} {100}},\ \bibinfo
  {pages} {224413} (\bibinfo {year} {2019})}\BibitemShut {NoStop}%
\bibitem [{\citenamefont {Anisimov}\ \emph {et~al.}(2019)\citenamefont
  {Anisimov}, \citenamefont {Aust}, \citenamefont {Khaliullin},\ and\
  \citenamefont {Daghofer}}]{Khaliullin2019_prl}%
  \BibitemOpen
  \bibfield  {author} {\bibinfo {author} {\bibfnamefont {P.~S.}\ \bibnamefont
  {Anisimov}}, \bibinfo {author} {\bibfnamefont {F.}~\bibnamefont {Aust}},
  \bibinfo {author} {\bibfnamefont {G.}~\bibnamefont {Khaliullin}},\ and\
  \bibinfo {author} {\bibfnamefont {M.}~\bibnamefont {Daghofer}},\ }\href
  {https://doi.org/10.1103/PhysRevLett.122.177201} {\bibfield  {journal}
  {\bibinfo  {journal} {Phys. Rev. Lett.}\ }\textbf {\bibinfo {volume} {122}},\
  \bibinfo {pages} {177201} (\bibinfo {year} {2019})}\BibitemShut {NoStop}%
\bibitem [{\citenamefont {Chaloupka}\ \emph {et~al.}(2010)\citenamefont
  {Chaloupka}, \citenamefont {Jackeli},\ and\ \citenamefont
  {Khaliullin}}]{Khaliullin_2010}%
  \BibitemOpen
  \bibfield  {author} {\bibinfo {author} {\bibfnamefont {J.~c.~v.}\
  \bibnamefont {Chaloupka}}, \bibinfo {author} {\bibfnamefont {G.}~\bibnamefont
  {Jackeli}},\ and\ \bibinfo {author} {\bibfnamefont {G.}~\bibnamefont
  {Khaliullin}},\ }\href {https://doi.org/10.1103/PhysRevLett.105.027204}
  {\bibfield  {journal} {\bibinfo  {journal} {Phys. Rev. Lett.}\ }\textbf
  {\bibinfo {volume} {105}},\ \bibinfo {pages} {027204} (\bibinfo {year}
  {2010})}\BibitemShut {NoStop}%
\bibitem [{\citenamefont {Takayama}\ \emph {et~al.}(2022)\citenamefont
  {Takayama}, \citenamefont {Blankenhorn}, \citenamefont {Bertinshaw},
  \citenamefont {Haskel}, \citenamefont {Bogdanov}, \citenamefont {Kitagawa},
  \citenamefont {Yaresko}, \citenamefont {Krajewska}, \citenamefont {Bette},
  \citenamefont {McNally}, \citenamefont {Gibbs}, \citenamefont {Matsumoto},
  \citenamefont {Sari}, \citenamefont {Watanabe}, \citenamefont {Fabbris},
  \citenamefont {Bi}, \citenamefont {Larkin}, \citenamefont {Rabinovich},
  \citenamefont {Boris}, \citenamefont {Ishii}, \citenamefont {Yamaoka},
  \citenamefont {Irifune}, \citenamefont {Bewley}, \citenamefont {Ridley},
  \citenamefont {Bull}, \citenamefont {Dinnebier}, \citenamefont {Keimer},\
  and\ \citenamefont {Takagi}}]{Takagi2022}%
  \BibitemOpen
  \bibfield  {author} {\bibinfo {author} {\bibfnamefont {T.}~\bibnamefont
  {Takayama}}, \bibinfo {author} {\bibfnamefont {M.}~\bibnamefont
  {Blankenhorn}}, \bibinfo {author} {\bibfnamefont {J.}~\bibnamefont
  {Bertinshaw}}, \bibinfo {author} {\bibfnamefont {D.}~\bibnamefont {Haskel}},
  \bibinfo {author} {\bibfnamefont {N.~A.}\ \bibnamefont {Bogdanov}}, \bibinfo
  {author} {\bibfnamefont {K.}~\bibnamefont {Kitagawa}}, \bibinfo {author}
  {\bibfnamefont {A.~N.}\ \bibnamefont {Yaresko}}, \bibinfo {author}
  {\bibfnamefont {A.}~\bibnamefont {Krajewska}}, \bibinfo {author}
  {\bibfnamefont {S.}~\bibnamefont {Bette}}, \bibinfo {author} {\bibfnamefont
  {G.}~\bibnamefont {McNally}}, \bibinfo {author} {\bibfnamefont {A.~S.}\
  \bibnamefont {Gibbs}}, \bibinfo {author} {\bibfnamefont {Y.}~\bibnamefont
  {Matsumoto}}, \bibinfo {author} {\bibfnamefont {D.~P.}\ \bibnamefont {Sari}},
  \bibinfo {author} {\bibfnamefont {I.}~\bibnamefont {Watanabe}}, \bibinfo
  {author} {\bibfnamefont {G.}~\bibnamefont {Fabbris}}, \bibinfo {author}
  {\bibfnamefont {W.}~\bibnamefont {Bi}}, \bibinfo {author} {\bibfnamefont
  {T.~I.}\ \bibnamefont {Larkin}}, \bibinfo {author} {\bibfnamefont {K.~S.}\
  \bibnamefont {Rabinovich}}, \bibinfo {author} {\bibfnamefont {A.~V.}\
  \bibnamefont {Boris}}, \bibinfo {author} {\bibfnamefont {H.}~\bibnamefont
  {Ishii}}, \bibinfo {author} {\bibfnamefont {H.}~\bibnamefont {Yamaoka}},
  \bibinfo {author} {\bibfnamefont {T.}~\bibnamefont {Irifune}}, \bibinfo
  {author} {\bibfnamefont {R.}~\bibnamefont {Bewley}}, \bibinfo {author}
  {\bibfnamefont {C.~J.}\ \bibnamefont {Ridley}}, \bibinfo {author}
  {\bibfnamefont {C.~L.}\ \bibnamefont {Bull}}, \bibinfo {author}
  {\bibfnamefont {R.}~\bibnamefont {Dinnebier}}, \bibinfo {author}
  {\bibfnamefont {B.}~\bibnamefont {Keimer}},\ and\ \bibinfo {author}
  {\bibfnamefont {H.}~\bibnamefont {Takagi}},\ }\href
  {https://doi.org/10.1103/PhysRevResearch.4.043079} {\bibfield  {journal}
  {\bibinfo  {journal} {Phys. Rev. Res.}\ }\textbf {\bibinfo {volume} {4}},\
  \bibinfo {pages} {043079} (\bibinfo {year} {2022})}\BibitemShut {NoStop}%
\bibitem [{\citenamefont {Williams}\ \emph {et~al.}(2016)\citenamefont
  {Williams}, \citenamefont {Johnson}, \citenamefont {Freund}, \citenamefont
  {Choi}, \citenamefont {Jesche}, \citenamefont {Kimchi}, \citenamefont
  {Manni}, \citenamefont {Bombardi}, \citenamefont {Manuel}, \citenamefont
  {Gegenwart},\ and\ \citenamefont {Coldea}}]{Coldea2016}%
  \BibitemOpen
  \bibfield  {author} {\bibinfo {author} {\bibfnamefont {S.~C.}\ \bibnamefont
  {Williams}}, \bibinfo {author} {\bibfnamefont {R.~D.}\ \bibnamefont
  {Johnson}}, \bibinfo {author} {\bibfnamefont {F.}~\bibnamefont {Freund}},
  \bibinfo {author} {\bibfnamefont {S.}~\bibnamefont {Choi}}, \bibinfo {author}
  {\bibfnamefont {A.}~\bibnamefont {Jesche}}, \bibinfo {author} {\bibfnamefont
  {I.}~\bibnamefont {Kimchi}}, \bibinfo {author} {\bibfnamefont
  {S.}~\bibnamefont {Manni}}, \bibinfo {author} {\bibfnamefont
  {A.}~\bibnamefont {Bombardi}}, \bibinfo {author} {\bibfnamefont
  {P.}~\bibnamefont {Manuel}}, \bibinfo {author} {\bibfnamefont
  {P.}~\bibnamefont {Gegenwart}},\ and\ \bibinfo {author} {\bibfnamefont
  {R.}~\bibnamefont {Coldea}},\ }\href
  {https://doi.org/10.1103/PhysRevB.93.195158} {\bibfield  {journal} {\bibinfo
  {journal} {Phys. Rev. B}\ }\textbf {\bibinfo {volume} {93}},\ \bibinfo
  {pages} {195158} (\bibinfo {year} {2016})}\BibitemShut {NoStop}%
\bibitem [{\citenamefont {Takagi}\ \emph {et~al.}(2019)\citenamefont {Takagi},
  \citenamefont {Takayama}, \citenamefont {Jackeli}, \citenamefont
  {Khaliullin},\ and\ \citenamefont {Nagler}}]{Takagi2019}%
  \BibitemOpen
  \bibfield  {author} {\bibinfo {author} {\bibfnamefont {H.}~\bibnamefont
  {Takagi}}, \bibinfo {author} {\bibfnamefont {T.}~\bibnamefont {Takayama}},
  \bibinfo {author} {\bibfnamefont {G.}~\bibnamefont {Jackeli}}, \bibinfo
  {author} {\bibfnamefont {G.}~\bibnamefont {Khaliullin}},\ and\ \bibinfo
  {author} {\bibfnamefont {S.~E.}\ \bibnamefont {Nagler}},\ }\href
  {https://doi.org/10.1038/s42254-019-0038-2} {\bibfield  {journal} {\bibinfo
  {journal} {Nature Reviews Physics}\ }\textbf {\bibinfo {volume} {1}},\
  \bibinfo {pages} {264} (\bibinfo {year} {2019})}\BibitemShut {NoStop}%
\bibitem [{\citenamefont {Singh}\ and\ \citenamefont
  {Gegenwart}(2010)}]{Yogesh2010}%
  \BibitemOpen
  \bibfield  {author} {\bibinfo {author} {\bibfnamefont {Y.}~\bibnamefont
  {Singh}}\ and\ \bibinfo {author} {\bibfnamefont {P.}~\bibnamefont
  {Gegenwart}},\ }\href {https://doi.org/10.1103/PhysRevB.82.064412} {\bibfield
   {journal} {\bibinfo  {journal} {Phys. Rev. B}\ }\textbf {\bibinfo {volume}
  {82}},\ \bibinfo {pages} {064412} (\bibinfo {year} {2010})}\BibitemShut
  {NoStop}%
\bibitem [{\citenamefont {Freund}\ \emph {et~al.}(2016)\citenamefont {Freund},
  \citenamefont {Williams}, \citenamefont {Johnson}, \citenamefont {Coldea},
  \citenamefont {Gegenwart},\ and\ \citenamefont {Jesche}}]{Freund2016}%
  \BibitemOpen
  \bibfield  {author} {\bibinfo {author} {\bibfnamefont {F.}~\bibnamefont
  {Freund}}, \bibinfo {author} {\bibfnamefont {S.~C.}\ \bibnamefont
  {Williams}}, \bibinfo {author} {\bibfnamefont {R.~D.}\ \bibnamefont
  {Johnson}}, \bibinfo {author} {\bibfnamefont {R.}~\bibnamefont {Coldea}},
  \bibinfo {author} {\bibfnamefont {P.}~\bibnamefont {Gegenwart}},\ and\
  \bibinfo {author} {\bibfnamefont {A.}~\bibnamefont {Jesche}},\ }\href
  {https://doi.org/10.1038/srep35362} {\bibfield  {journal} {\bibinfo
  {journal} {Scientific Reports}\ }\textbf {\bibinfo {volume} {6}},\ \bibinfo
  {pages} {35362} (\bibinfo {year} {2016})}\BibitemShut {NoStop}%
\bibitem [{\citenamefont {Choi}\ \emph {et~al.}(2012)\citenamefont {Choi},
  \citenamefont {Coldea}, \citenamefont {Kolmogorov}, \citenamefont
  {Lancaster}, \citenamefont {Mazin}, \citenamefont {Blundell}, \citenamefont
  {Radaelli}, \citenamefont {Singh}, \citenamefont {Gegenwart}, \citenamefont
  {Choi}, \citenamefont {Cheong}, \citenamefont {Baker}, \citenamefont
  {Stock},\ and\ \citenamefont {Taylor}}]{Choi2012}%
  \BibitemOpen
  \bibfield  {author} {\bibinfo {author} {\bibfnamefont {S.~K.}\ \bibnamefont
  {Choi}}, \bibinfo {author} {\bibfnamefont {R.}~\bibnamefont {Coldea}},
  \bibinfo {author} {\bibfnamefont {A.~N.}\ \bibnamefont {Kolmogorov}},
  \bibinfo {author} {\bibfnamefont {T.}~\bibnamefont {Lancaster}}, \bibinfo
  {author} {\bibfnamefont {I.~I.}\ \bibnamefont {Mazin}}, \bibinfo {author}
  {\bibfnamefont {S.~J.}\ \bibnamefont {Blundell}}, \bibinfo {author}
  {\bibfnamefont {P.~G.}\ \bibnamefont {Radaelli}}, \bibinfo {author}
  {\bibfnamefont {Y.}~\bibnamefont {Singh}}, \bibinfo {author} {\bibfnamefont
  {P.}~\bibnamefont {Gegenwart}}, \bibinfo {author} {\bibfnamefont {K.~R.}\
  \bibnamefont {Choi}}, \bibinfo {author} {\bibfnamefont {S.-W.}\ \bibnamefont
  {Cheong}}, \bibinfo {author} {\bibfnamefont {P.~J.}\ \bibnamefont {Baker}},
  \bibinfo {author} {\bibfnamefont {C.}~\bibnamefont {Stock}},\ and\ \bibinfo
  {author} {\bibfnamefont {J.}~\bibnamefont {Taylor}},\ }\href
  {https://doi.org/10.1103/PhysRevLett.108.127204} {\bibfield  {journal}
  {\bibinfo  {journal} {Phys. Rev. Lett.}\ }\textbf {\bibinfo {volume} {108}},\
  \bibinfo {pages} {127204} (\bibinfo {year} {2012})}\BibitemShut {NoStop}%
\bibitem [{\citenamefont {Ghosh}\ \emph {et~al.}(2021)\citenamefont {Ghosh},
  \citenamefont {Shekhter}, \citenamefont {Jerzembeck}, \citenamefont
  {Kikugawa}, \citenamefont {Sokolov}, \citenamefont {Brando}, \citenamefont
  {Mackenzie}, \citenamefont {Hicks},\ and\ \citenamefont
  {Ramshaw}}]{Ghosh2021}%
  \BibitemOpen
  \bibfield  {author} {\bibinfo {author} {\bibfnamefont {S.}~\bibnamefont
  {Ghosh}}, \bibinfo {author} {\bibfnamefont {A.}~\bibnamefont {Shekhter}},
  \bibinfo {author} {\bibfnamefont {F.}~\bibnamefont {Jerzembeck}}, \bibinfo
  {author} {\bibfnamefont {N.}~\bibnamefont {Kikugawa}}, \bibinfo {author}
  {\bibfnamefont {D.~A.}\ \bibnamefont {Sokolov}}, \bibinfo {author}
  {\bibfnamefont {M.}~\bibnamefont {Brando}}, \bibinfo {author} {\bibfnamefont
  {A.~P.}\ \bibnamefont {Mackenzie}}, \bibinfo {author} {\bibfnamefont {C.~W.}\
  \bibnamefont {Hicks}},\ and\ \bibinfo {author} {\bibfnamefont {B.~J.}\
  \bibnamefont {Ramshaw}},\ }\href {https://doi.org/10.1038/s41567-020-1032-4}
  {\bibfield  {journal} {\bibinfo  {journal} {Nature Physics}\ }\textbf
  {\bibinfo {volume} {17}},\ \bibinfo {pages} {199} (\bibinfo {year}
  {2021})}\BibitemShut {NoStop}%
\bibitem [{\citenamefont {Kumar}\ \emph {et~al.}(2016)\citenamefont {Kumar},
  \citenamefont {Sheptyakov}, \citenamefont {Khuntia}, \citenamefont {Rolfs},
  \citenamefont {Freeman}, \citenamefont {R\o{}nnow}, \citenamefont {Dey},
  \citenamefont {Baenitz},\ and\ \citenamefont {Mahajan}}]{AVM2016}%
  \BibitemOpen
  \bibfield  {author} {\bibinfo {author} {\bibfnamefont {R.}~\bibnamefont
  {Kumar}}, \bibinfo {author} {\bibfnamefont {D.}~\bibnamefont {Sheptyakov}},
  \bibinfo {author} {\bibfnamefont {P.}~\bibnamefont {Khuntia}}, \bibinfo
  {author} {\bibfnamefont {K.}~\bibnamefont {Rolfs}}, \bibinfo {author}
  {\bibfnamefont {P.~G.}\ \bibnamefont {Freeman}}, \bibinfo {author}
  {\bibfnamefont {H.~M.}\ \bibnamefont {R\o{}nnow}}, \bibinfo {author}
  {\bibfnamefont {T.}~\bibnamefont {Dey}}, \bibinfo {author} {\bibfnamefont
  {M.}~\bibnamefont {Baenitz}},\ and\ \bibinfo {author} {\bibfnamefont {A.~V.}\
  \bibnamefont {Mahajan}},\ }\href {https://doi.org/10.1103/PhysRevB.94.174410}
  {\bibfield  {journal} {\bibinfo  {journal} {Phys. Rev. B}\ }\textbf {\bibinfo
  {volume} {94}},\ \bibinfo {pages} {174410} (\bibinfo {year}
  {2016})}\BibitemShut {NoStop}%
\bibitem [{\citenamefont {Kumar}\ \emph
  {et~al.}(2019{\natexlab{a}})\citenamefont {Kumar}, \citenamefont {Dey},
  \citenamefont {Ette}, \citenamefont {Ramesha}, \citenamefont {Chakraborty},
  \citenamefont {Dasgupta}, \citenamefont {Orain}, \citenamefont {Baines},
  \citenamefont {T\'oth}, \citenamefont {Shahee}, \citenamefont {Kundu},
  \citenamefont {Prinz-Zwick}, \citenamefont {Gippius}, \citenamefont
  {B\"uttgen}, \citenamefont {Gegenwart},\ and\ \citenamefont
  {Mahajan}}]{RKumar2019}%
  \BibitemOpen
  \bibfield  {author} {\bibinfo {author} {\bibfnamefont {R.}~\bibnamefont
  {Kumar}}, \bibinfo {author} {\bibfnamefont {T.}~\bibnamefont {Dey}}, \bibinfo
  {author} {\bibfnamefont {P.~M.}\ \bibnamefont {Ette}}, \bibinfo {author}
  {\bibfnamefont {K.}~\bibnamefont {Ramesha}}, \bibinfo {author} {\bibfnamefont
  {A.}~\bibnamefont {Chakraborty}}, \bibinfo {author} {\bibfnamefont
  {I.}~\bibnamefont {Dasgupta}}, \bibinfo {author} {\bibfnamefont {J.~C.}\
  \bibnamefont {Orain}}, \bibinfo {author} {\bibfnamefont {C.}~\bibnamefont
  {Baines}}, \bibinfo {author} {\bibfnamefont {S.}~\bibnamefont {T\'oth}},
  \bibinfo {author} {\bibfnamefont {A.}~\bibnamefont {Shahee}}, \bibinfo
  {author} {\bibfnamefont {S.}~\bibnamefont {Kundu}}, \bibinfo {author}
  {\bibfnamefont {M.}~\bibnamefont {Prinz-Zwick}}, \bibinfo {author}
  {\bibfnamefont {A.~A.}\ \bibnamefont {Gippius}}, \bibinfo {author}
  {\bibfnamefont {N.}~\bibnamefont {B\"uttgen}}, \bibinfo {author}
  {\bibfnamefont {P.}~\bibnamefont {Gegenwart}},\ and\ \bibinfo {author}
  {\bibfnamefont {A.~V.}\ \bibnamefont {Mahajan}},\ }\href
  {https://doi.org/10.1103/PhysRevB.99.054417} {\bibfield  {journal} {\bibinfo
  {journal} {Phys. Rev. B}\ }\textbf {\bibinfo {volume} {99}},\ \bibinfo
  {pages} {054417} (\bibinfo {year} {2019}{\natexlab{a}})}\BibitemShut
  {NoStop}%
\bibitem [{\citenamefont {Kumar}\ \emph
  {et~al.}(2019{\natexlab{b}})\citenamefont {Kumar}, \citenamefont {Dey},
  \citenamefont {Ette}, \citenamefont {Ramesha}, \citenamefont {Chakraborty},
  \citenamefont {Dasgupta}, \citenamefont {Eremina}, \citenamefont {T\'oth},
  \citenamefont {Shahee}, \citenamefont {Kundu}, \citenamefont {Prinz-Zwick},
  \citenamefont {Gippius}, \citenamefont {von Nidda}, \citenamefont
  {B\"uttgen}, \citenamefont {Gegenwart},\ and\ \citenamefont
  {Mahajan}}]{RKumar2019ALMO}%
  \BibitemOpen
  \bibfield  {author} {\bibinfo {author} {\bibfnamefont {R.}~\bibnamefont
  {Kumar}}, \bibinfo {author} {\bibfnamefont {T.}~\bibnamefont {Dey}}, \bibinfo
  {author} {\bibfnamefont {P.~M.}\ \bibnamefont {Ette}}, \bibinfo {author}
  {\bibfnamefont {K.}~\bibnamefont {Ramesha}}, \bibinfo {author} {\bibfnamefont
  {A.}~\bibnamefont {Chakraborty}}, \bibinfo {author} {\bibfnamefont
  {I.}~\bibnamefont {Dasgupta}}, \bibinfo {author} {\bibfnamefont
  {R.}~\bibnamefont {Eremina}}, \bibinfo {author} {\bibfnamefont
  {S.}~\bibnamefont {T\'oth}}, \bibinfo {author} {\bibfnamefont
  {A.}~\bibnamefont {Shahee}}, \bibinfo {author} {\bibfnamefont
  {S.}~\bibnamefont {Kundu}}, \bibinfo {author} {\bibfnamefont
  {M.}~\bibnamefont {Prinz-Zwick}}, \bibinfo {author} {\bibfnamefont {A.~A.}\
  \bibnamefont {Gippius}}, \bibinfo {author} {\bibfnamefont {H.~A.~K.}\
  \bibnamefont {von Nidda}}, \bibinfo {author} {\bibfnamefont {N.}~\bibnamefont
  {B\"uttgen}}, \bibinfo {author} {\bibfnamefont {P.}~\bibnamefont
  {Gegenwart}},\ and\ \bibinfo {author} {\bibfnamefont {A.~V.}\ \bibnamefont
  {Mahajan}},\ }\href {https://doi.org/10.1103/PhysRevB.99.144429} {\bibfield
  {journal} {\bibinfo  {journal} {Phys. Rev. B}\ }\textbf {\bibinfo {volume}
  {99}},\ \bibinfo {pages} {144429} (\bibinfo {year}
  {2019}{\natexlab{b}})}\BibitemShut {NoStop}%
\bibitem [{\citenamefont {Flint}\ and\ \citenamefont {Lee}(2013)}]{Lee2013}%
  \BibitemOpen
  \bibfield  {author} {\bibinfo {author} {\bibfnamefont {R.}~\bibnamefont
  {Flint}}\ and\ \bibinfo {author} {\bibfnamefont {P.~A.}\ \bibnamefont
  {Lee}},\ }\href {https://doi.org/10.1103/PhysRevLett.111.217201} {\bibfield
  {journal} {\bibinfo  {journal} {Phys. Rev. Lett.}\ }\textbf {\bibinfo
  {volume} {111}},\ \bibinfo {pages} {217201} (\bibinfo {year}
  {2013})}\BibitemShut {NoStop}%
\bibitem [{\citenamefont {Svoboda}\ \emph {et~al.}(2017)\citenamefont
  {Svoboda}, \citenamefont {Randeria},\ and\ \citenamefont
  {Trivedi}}]{Trivedi2017}%
  \BibitemOpen
  \bibfield  {author} {\bibinfo {author} {\bibfnamefont {C.}~\bibnamefont
  {Svoboda}}, \bibinfo {author} {\bibfnamefont {M.}~\bibnamefont {Randeria}},\
  and\ \bibinfo {author} {\bibfnamefont {N.}~\bibnamefont {Trivedi}},\ }\href
  {https://doi.org/10.1103/PhysRevB.95.014409} {\bibfield  {journal} {\bibinfo
  {journal} {Phys. Rev. B}\ }\textbf {\bibinfo {volume} {95}},\ \bibinfo
  {pages} {014409} (\bibinfo {year} {2017})}\BibitemShut {NoStop}%
\bibitem [{\citenamefont {Kitagawa}\ \emph {et~al.}(2018)\citenamefont
  {Kitagawa}, \citenamefont {Takayama}, \citenamefont {Matsumoto},
  \citenamefont {Kato}, \citenamefont {Takano}, \citenamefont {Kishimoto},
  \citenamefont {Bette}, \citenamefont {Dinnebier}, \citenamefont {Jackeli},\
  and\ \citenamefont {Takagi}}]{Kitagawa2018}%
  \BibitemOpen
  \bibfield  {author} {\bibinfo {author} {\bibfnamefont {K.}~\bibnamefont
  {Kitagawa}}, \bibinfo {author} {\bibfnamefont {T.}~\bibnamefont {Takayama}},
  \bibinfo {author} {\bibfnamefont {Y.}~\bibnamefont {Matsumoto}}, \bibinfo
  {author} {\bibfnamefont {A.}~\bibnamefont {Kato}}, \bibinfo {author}
  {\bibfnamefont {R.}~\bibnamefont {Takano}}, \bibinfo {author} {\bibfnamefont
  {Y.}~\bibnamefont {Kishimoto}}, \bibinfo {author} {\bibfnamefont
  {S.}~\bibnamefont {Bette}}, \bibinfo {author} {\bibfnamefont
  {R.}~\bibnamefont {Dinnebier}}, \bibinfo {author} {\bibfnamefont
  {G.}~\bibnamefont {Jackeli}},\ and\ \bibinfo {author} {\bibfnamefont
  {H.}~\bibnamefont {Takagi}},\ }\href {https://doi.org/10.1038/nature25482}
  {\bibfield  {journal} {\bibinfo  {journal} {Nature}\ }\textbf {\bibinfo
  {volume} {554}},\ \bibinfo {pages} {341} (\bibinfo {year}
  {2018})}\BibitemShut {NoStop}%
\bibitem [{\citenamefont {Takahashi}\ \emph {et~al.}(2019)\citenamefont
  {Takahashi}, \citenamefont {Wang}, \citenamefont {Arsenault}, \citenamefont
  {Imai}, \citenamefont {Abramchuk}, \citenamefont {Tafti},\ and\ \citenamefont
  {Singer}}]{Singer2019}%
  \BibitemOpen
  \bibfield  {author} {\bibinfo {author} {\bibfnamefont {S.~K.}\ \bibnamefont
  {Takahashi}}, \bibinfo {author} {\bibfnamefont {J.}~\bibnamefont {Wang}},
  \bibinfo {author} {\bibfnamefont {A.}~\bibnamefont {Arsenault}}, \bibinfo
  {author} {\bibfnamefont {T.}~\bibnamefont {Imai}}, \bibinfo {author}
  {\bibfnamefont {M.}~\bibnamefont {Abramchuk}}, \bibinfo {author}
  {\bibfnamefont {F.}~\bibnamefont {Tafti}},\ and\ \bibinfo {author}
  {\bibfnamefont {P.~M.}\ \bibnamefont {Singer}},\ }\href
  {https://doi.org/10.1103/PhysRevX.9.031047} {\bibfield  {journal} {\bibinfo
  {journal} {Phys. Rev. X}\ }\textbf {\bibinfo {volume} {9}},\ \bibinfo {pages}
  {031047} (\bibinfo {year} {2019})}\BibitemShut {NoStop}%
\bibitem [{SM()}]{SM}%
  \BibitemOpen
  \href@noop {} {}\bibinfo {note} {For details see Supplemental Material at URL
  , which includes
  Refs.~\onlinecite{Kittel2005,SKundu2020_YCTO,Bahrami2021_prb,Huang2020}}\BibitemShut
  {NoStop}%
\bibitem [{Note1()}]{Note1}%
  \BibitemOpen
  \bibinfo {note} {Low-$T$ Curie-tail is from about 3\% S=1/2 defects
  impurity.}\BibitemShut {Stop}%
\bibitem [{Note2()}]{Note2}%
  \BibitemOpen
  \bibinfo {note} {Dynamical spin susceptibility is a measure of the response
  of a material's magnetic moment to an applied magnetic field. Specifically,
  it is the Fourier transform of the auto-correlation function of the spin
  fluctuations in the material.}\BibitemShut {Stop}%
\bibitem [{\citenamefont {Pustogow}\ \emph {et~al.}(2020)\citenamefont
  {Pustogow}, \citenamefont {Le}, \citenamefont {Wang}, \citenamefont {Luo},
  \citenamefont {Gati}, \citenamefont {Schubert}, \citenamefont {Lang},\ and\
  \citenamefont {Brown}}]{Fermi-Liquid-1}%
  \BibitemOpen
  \bibfield  {author} {\bibinfo {author} {\bibfnamefont {A.}~\bibnamefont
  {Pustogow}}, \bibinfo {author} {\bibfnamefont {T.}~\bibnamefont {Le}},
  \bibinfo {author} {\bibfnamefont {H.-H.}\ \bibnamefont {Wang}}, \bibinfo
  {author} {\bibfnamefont {Y.}~\bibnamefont {Luo}}, \bibinfo {author}
  {\bibfnamefont {E.}~\bibnamefont {Gati}}, \bibinfo {author} {\bibfnamefont
  {H.}~\bibnamefont {Schubert}}, \bibinfo {author} {\bibfnamefont
  {M.}~\bibnamefont {Lang}},\ and\ \bibinfo {author} {\bibfnamefont {S.~E.}\
  \bibnamefont {Brown}},\ }\href {https://doi.org/10.1103/PhysRevB.101.140401}
  {\bibfield  {journal} {\bibinfo  {journal} {Phys. Rev. B}\ }\textbf {\bibinfo
  {volume} {101}},\ \bibinfo {pages} {140401} (\bibinfo {year}
  {2020})}\BibitemShut {NoStop}%
\bibitem [{\citenamefont {Tewari}\ and\ \citenamefont
  {Ruvalds}(1996)}]{Fermi1996}%
  \BibitemOpen
  \bibfield  {author} {\bibinfo {author} {\bibfnamefont {S.}~\bibnamefont
  {Tewari}}\ and\ \bibinfo {author} {\bibfnamefont {J.}~\bibnamefont
  {Ruvalds}},\ }\href {https://doi.org/10.1103/PhysRevB.53.5696} {\bibfield
  {journal} {\bibinfo  {journal} {Phys. Rev. B}\ }\textbf {\bibinfo {volume}
  {53}},\ \bibinfo {pages} {5696} (\bibinfo {year} {1996})}\BibitemShut
  {NoStop}%
\bibitem [{\citenamefont {Li}\ \emph {et~al.}(2016)\citenamefont {Li},
  \citenamefont {Adroja}, \citenamefont {Biswas}, \citenamefont {Baker},
  \citenamefont {Zhang}, \citenamefont {Liu}, \citenamefont {Tsirlin},
  \citenamefont {Gegenwart},\ and\ \citenamefont {Zhang}}]{YMGO}%
  \BibitemOpen
  \bibfield  {author} {\bibinfo {author} {\bibfnamefont {Y.}~\bibnamefont
  {Li}}, \bibinfo {author} {\bibfnamefont {D.}~\bibnamefont {Adroja}}, \bibinfo
  {author} {\bibfnamefont {P.~K.}\ \bibnamefont {Biswas}}, \bibinfo {author}
  {\bibfnamefont {P.~J.}\ \bibnamefont {Baker}}, \bibinfo {author}
  {\bibfnamefont {Q.}~\bibnamefont {Zhang}}, \bibinfo {author} {\bibfnamefont
  {J.}~\bibnamefont {Liu}}, \bibinfo {author} {\bibfnamefont {A.~A.}\
  \bibnamefont {Tsirlin}}, \bibinfo {author} {\bibfnamefont {P.}~\bibnamefont
  {Gegenwart}},\ and\ \bibinfo {author} {\bibfnamefont {Q.}~\bibnamefont
  {Zhang}},\ }\href {https://doi.org/10.1103/PhysRevLett.117.097201} {\bibfield
   {journal} {\bibinfo  {journal} {Phys. Rev. Lett.}\ }\textbf {\bibinfo
  {volume} {117}},\ \bibinfo {pages} {097201} (\bibinfo {year}
  {2016})}\BibitemShut {NoStop}%
\bibitem [{\citenamefont {Kundu}\ \emph
  {et~al.}(2020{\natexlab{a}})\citenamefont {Kundu}, \citenamefont {Shahee},
  \citenamefont {Chakraborty}, \citenamefont {Ranjith}, \citenamefont {Koo},
  \citenamefont {Sichelschmidt}, \citenamefont {Telling}, \citenamefont
  {Biswas}, \citenamefont {Baenitz}, \citenamefont {Dasgupta}, \citenamefont
  {Pujari},\ and\ \citenamefont {Mahajan}}]{SKundu2020}%
  \BibitemOpen
  \bibfield  {author} {\bibinfo {author} {\bibfnamefont {S.}~\bibnamefont
  {Kundu}}, \bibinfo {author} {\bibfnamefont {A.}~\bibnamefont {Shahee}},
  \bibinfo {author} {\bibfnamefont {A.}~\bibnamefont {Chakraborty}}, \bibinfo
  {author} {\bibfnamefont {K.~M.}\ \bibnamefont {Ranjith}}, \bibinfo {author}
  {\bibfnamefont {B.}~\bibnamefont {Koo}}, \bibinfo {author} {\bibfnamefont
  {J.}~\bibnamefont {Sichelschmidt}}, \bibinfo {author} {\bibfnamefont
  {M.~T.~F.}\ \bibnamefont {Telling}}, \bibinfo {author} {\bibfnamefont
  {P.~K.}\ \bibnamefont {Biswas}}, \bibinfo {author} {\bibfnamefont
  {M.}~\bibnamefont {Baenitz}}, \bibinfo {author} {\bibfnamefont
  {I.}~\bibnamefont {Dasgupta}}, \bibinfo {author} {\bibfnamefont
  {S.}~\bibnamefont {Pujari}},\ and\ \bibinfo {author} {\bibfnamefont {A.~V.}\
  \bibnamefont {Mahajan}},\ }\href
  {https://doi.org/10.1103/PhysRevLett.125.267202} {\bibfield  {journal}
  {\bibinfo  {journal} {Phys. Rev. Lett.}\ }\textbf {\bibinfo {volume} {125}},\
  \bibinfo {pages} {267202} (\bibinfo {year} {2020}{\natexlab{a}})}\BibitemShut
  {NoStop}%
\bibitem [{\citenamefont {Kittel}\ and\ \citenamefont
  {McEuen}(2005)}]{Kittel2005}%
  \BibitemOpen
  \bibfield  {author} {\bibinfo {author} {\bibfnamefont {C.}~\bibnamefont
  {Kittel}}\ and\ \bibinfo {author} {\bibfnamefont {P.}~\bibnamefont
  {McEuen}},\ }\href@noop {} {\emph {\bibinfo {title} {Introduction to solid
  state physics}}},\ \bibinfo {edition} {8th}\ ed.\ (\bibinfo  {publisher}
  {John Wiley \& Sons, Inc},\ \bibinfo {year} {2005})\BibitemShut {NoStop}%
\bibitem [{\citenamefont {Kundu}\ \emph
  {et~al.}(2020{\natexlab{b}})\citenamefont {Kundu}, \citenamefont {Hossain},
  \citenamefont {S.}, \citenamefont {Das}, \citenamefont {Baenitz},
  \citenamefont {Baker}, \citenamefont {Orain}, \citenamefont {Joshi},
  \citenamefont {Mathieu}, \citenamefont {Mahadevan}, \citenamefont {Pujari},
  \citenamefont {Bhattacharjee}, \citenamefont {Mahajan},\ and\ \citenamefont
  {Sarma}}]{SKundu2020_YCTO}%
  \BibitemOpen
  \bibfield  {author} {\bibinfo {author} {\bibfnamefont {S.}~\bibnamefont
  {Kundu}}, \bibinfo {author} {\bibfnamefont {A.}~\bibnamefont {Hossain}},
  \bibinfo {author} {\bibfnamefont {P.~K.}\ \bibnamefont {S.}}, \bibinfo
  {author} {\bibfnamefont {R.}~\bibnamefont {Das}}, \bibinfo {author}
  {\bibfnamefont {M.}~\bibnamefont {Baenitz}}, \bibinfo {author} {\bibfnamefont
  {P.~J.}\ \bibnamefont {Baker}}, \bibinfo {author} {\bibfnamefont {J.-C.}\
  \bibnamefont {Orain}}, \bibinfo {author} {\bibfnamefont {D.~C.}\ \bibnamefont
  {Joshi}}, \bibinfo {author} {\bibfnamefont {R.}~\bibnamefont {Mathieu}},
  \bibinfo {author} {\bibfnamefont {P.}~\bibnamefont {Mahadevan}}, \bibinfo
  {author} {\bibfnamefont {S.}~\bibnamefont {Pujari}}, \bibinfo {author}
  {\bibfnamefont {S.}~\bibnamefont {Bhattacharjee}}, \bibinfo {author}
  {\bibfnamefont {A.~V.}\ \bibnamefont {Mahajan}},\ and\ \bibinfo {author}
  {\bibfnamefont {D.~D.}\ \bibnamefont {Sarma}},\ }\href
  {https://doi.org/10.1103/PhysRevLett.125.117206} {\bibfield  {journal}
  {\bibinfo  {journal} {Phys. Rev. Lett.}\ }\textbf {\bibinfo {volume} {125}},\
  \bibinfo {pages} {117206} (\bibinfo {year} {2020}{\natexlab{b}})}\BibitemShut
  {NoStop}%
\bibitem [{\citenamefont {Bahrami}\ \emph {et~al.}(2021)\citenamefont
  {Bahrami}, \citenamefont {Kenney}, \citenamefont {Wang}, \citenamefont
  {Berlie}, \citenamefont {Lebedev}, \citenamefont {Graf},\ and\ \citenamefont
  {Tafti}}]{Bahrami2021_prb}%
  \BibitemOpen
  \bibfield  {author} {\bibinfo {author} {\bibfnamefont {F.}~\bibnamefont
  {Bahrami}}, \bibinfo {author} {\bibfnamefont {E.~M.}\ \bibnamefont {Kenney}},
  \bibinfo {author} {\bibfnamefont {C.}~\bibnamefont {Wang}}, \bibinfo {author}
  {\bibfnamefont {A.}~\bibnamefont {Berlie}}, \bibinfo {author} {\bibfnamefont
  {O.~I.}\ \bibnamefont {Lebedev}}, \bibinfo {author} {\bibfnamefont {M.~J.}\
  \bibnamefont {Graf}},\ and\ \bibinfo {author} {\bibfnamefont
  {F.}~\bibnamefont {Tafti}},\ }\href
  {https://doi.org/10.1103/PhysRevB.103.094427} {\bibfield  {journal} {\bibinfo
   {journal} {Phys. Rev. B}\ }\textbf {\bibinfo {volume} {103}},\ \bibinfo
  {pages} {094427} (\bibinfo {year} {2021})}\BibitemShut {NoStop}%
\bibitem [{\citenamefont {Li}\ \emph {et~al.}(2020)\citenamefont {Li},
  \citenamefont {Huang}, \citenamefont {Chen}, \citenamefont {Liu},
  \citenamefont {Pei}, \citenamefont {Wang}, \citenamefont {Wang},
  \citenamefont {Zhao}, \citenamefont {Yu}, \citenamefont {Wang}, \citenamefont
  {Ye}, \citenamefont {Mei},\ and\ \citenamefont {Huang}}]{Huang2020}%
  \BibitemOpen
  \bibfield  {author} {\bibinfo {author} {\bibfnamefont {G.}~\bibnamefont
  {Li}}, \bibinfo {author} {\bibfnamefont {L.-L.}\ \bibnamefont {Huang}},
  \bibinfo {author} {\bibfnamefont {X.}~\bibnamefont {Chen}}, \bibinfo {author}
  {\bibfnamefont {C.}~\bibnamefont {Liu}}, \bibinfo {author} {\bibfnamefont
  {S.}~\bibnamefont {Pei}}, \bibinfo {author} {\bibfnamefont {X.}~\bibnamefont
  {Wang}}, \bibinfo {author} {\bibfnamefont {S.}~\bibnamefont {Wang}}, \bibinfo
  {author} {\bibfnamefont {Y.}~\bibnamefont {Zhao}}, \bibinfo {author}
  {\bibfnamefont {D.}~\bibnamefont {Yu}}, \bibinfo {author} {\bibfnamefont
  {L.}~\bibnamefont {Wang}}, \bibinfo {author} {\bibfnamefont {F.}~\bibnamefont
  {Ye}}, \bibinfo {author} {\bibfnamefont {J.-W.}\ \bibnamefont {Mei}},\ and\
  \bibinfo {author} {\bibfnamefont {M.}~\bibnamefont {Huang}},\ }\href
  {https://doi.org/10.1103/PhysRevB.101.174436} {\bibfield  {journal} {\bibinfo
   {journal} {Phys. Rev. B}\ }\textbf {\bibinfo {volume} {101}},\ \bibinfo
  {pages} {174436} (\bibinfo {year} {2020})}\BibitemShut {NoStop}%
\end{thebibliography}%


\begin{thebibliography}{7}%
\makeatletter
\providecommand \@ifxundefined [1]{%
 \@ifx{#1\undefined}
}%
\providecommand \@ifnum [1]{%
 \ifnum #1\expandafter \@firstoftwo
 \else \expandafter \@secondoftwo
 \fi
}%
\providecommand \@ifx [1]{%
 \ifx #1\expandafter \@firstoftwo
 \else \expandafter \@secondoftwo
 \fi
}%
\providecommand \natexlab [1]{#1}%
\providecommand \enquote  [1]{``#1''}%
\providecommand \bibnamefont  [1]{#1}%
\providecommand \bibfnamefont [1]{#1}%
\providecommand \citenamefont [1]{#1}%
\providecommand \href@noop [0]{\@secondoftwo}%
\providecommand \href [0]{\begingroup \@sanitize@url \@href}%
\providecommand \@href[1]{\@@startlink{#1}\@@href}%
\providecommand \@@href[1]{\endgroup#1\@@endlink}%
\providecommand \@sanitize@url [0]{\catcode `\\12\catcode `\$12\catcode
  `\&12\catcode `\#12\catcode `\^12\catcode `\_12\catcode `\%12\relax}%
\providecommand \@@startlink[1]{}%
\providecommand \@@endlink[0]{}%
\providecommand \url  [0]{\begingroup\@sanitize@url \@url }%
\providecommand \@url [1]{\endgroup\@href {#1}{\urlprefix }}%
\providecommand \urlprefix  [0]{URL }%
\providecommand \Eprint [0]{\href }%
\providecommand \doibase [0]{https://doi.org/}%
\providecommand \selectlanguage [0]{\@gobble}%
\providecommand \bibinfo  [0]{\@secondoftwo}%
\providecommand \bibfield  [0]{\@secondoftwo}%
\providecommand \translation [1]{[#1]}%
\providecommand \BibitemOpen [0]{}%
\providecommand \bibitemStop [0]{}%
\providecommand \bibitemNoStop [0]{.\EOS\space}%
\providecommand \EOS [0]{\spacefactor3000\relax}%
\providecommand \BibitemShut  [1]{\csname bibitem#1\endcsname}%
\let\auto@bib@innerbib\@empty
\bibitem [{\citenamefont {Li}\ \emph {et~al.}(2020)\citenamefont {Li},
  \citenamefont {Huang}, \citenamefont {Chen}, \citenamefont {Liu},
  \citenamefont {Pei}, \citenamefont {Wang}, \citenamefont {Wang},
  \citenamefont {Zhao}, \citenamefont {Yu}, \citenamefont {Wang}, \citenamefont
  {Ye}, \citenamefont {Mei},\ and\ \citenamefont {Huang}}]{Huang2020}%
  \BibitemOpen
  \bibfield  {author} {\bibinfo {author} {\bibfnamefont {G.}~\bibnamefont
  {Li}}, \bibinfo {author} {\bibfnamefont {L.-L.}\ \bibnamefont {Huang}},
  \bibinfo {author} {\bibfnamefont {X.}~\bibnamefont {Chen}}, \bibinfo {author}
  {\bibfnamefont {C.}~\bibnamefont {Liu}}, \bibinfo {author} {\bibfnamefont
  {S.}~\bibnamefont {Pei}}, \bibinfo {author} {\bibfnamefont {X.}~\bibnamefont
  {Wang}}, \bibinfo {author} {\bibfnamefont {S.}~\bibnamefont {Wang}}, \bibinfo
  {author} {\bibfnamefont {Y.}~\bibnamefont {Zhao}}, \bibinfo {author}
  {\bibfnamefont {D.}~\bibnamefont {Yu}}, \bibinfo {author} {\bibfnamefont
  {L.}~\bibnamefont {Wang}}, \bibinfo {author} {\bibfnamefont {F.}~\bibnamefont
  {Ye}}, \bibinfo {author} {\bibfnamefont {J.-W.}\ \bibnamefont {Mei}},\ and\
  \bibinfo {author} {\bibfnamefont {M.}~\bibnamefont {Huang}},\ }\href
  {https://doi.org/10.1103/PhysRevB.101.174436} {\bibfield  {journal} {\bibinfo
   {journal} {Phys. Rev. B}\ }\textbf {\bibinfo {volume} {101}},\ \bibinfo
  {pages} {174436} (\bibinfo {year} {2020})}\BibitemShut {NoStop}%
\bibitem [{\citenamefont {Bahrami}\ \emph {et~al.}(2019)\citenamefont
  {Bahrami}, \citenamefont {Lafargue-Dit-Hauret}, \citenamefont {Lebedev},
  \citenamefont {Movshovich}, \citenamefont {Yang}, \citenamefont {Broido},
  \citenamefont {Rocquefelte},\ and\ \citenamefont {Tafti}}]{Bahrami_T_2019}%
  \BibitemOpen
  \bibfield  {author} {\bibinfo {author} {\bibfnamefont {F.}~\bibnamefont
  {Bahrami}}, \bibinfo {author} {\bibfnamefont {W.}~\bibnamefont
  {Lafargue-Dit-Hauret}}, \bibinfo {author} {\bibfnamefont {O.~I.}\
  \bibnamefont {Lebedev}}, \bibinfo {author} {\bibfnamefont {R.}~\bibnamefont
  {Movshovich}}, \bibinfo {author} {\bibfnamefont {H.-Y.}\ \bibnamefont
  {Yang}}, \bibinfo {author} {\bibfnamefont {D.}~\bibnamefont {Broido}},
  \bibinfo {author} {\bibfnamefont {X.}~\bibnamefont {Rocquefelte}},\ and\
  \bibinfo {author} {\bibfnamefont {F.}~\bibnamefont {Tafti}},\ }\href
  {https://doi.org/10.1103/PhysRevLett.123.237203} {\bibfield  {journal}
  {\bibinfo  {journal} {Phys. Rev. Lett.}\ }\textbf {\bibinfo {volume} {123}},\
  \bibinfo {pages} {237203} (\bibinfo {year} {2019})}\BibitemShut {NoStop}%
\bibitem [{\citenamefont {Kundu}\ \emph
  {et~al.}(2020{\natexlab{a}})\citenamefont {Kundu}, \citenamefont {Hossain},
  \citenamefont {S.}, \citenamefont {Das}, \citenamefont {Baenitz},
  \citenamefont {Baker}, \citenamefont {Orain}, \citenamefont {Joshi},
  \citenamefont {Mathieu}, \citenamefont {Mahadevan}, \citenamefont {Pujari},
  \citenamefont {Bhattacharjee}, \citenamefont {Mahajan},\ and\ \citenamefont
  {Sarma}}]{SKundu2020_YCTO}%
  \BibitemOpen
  \bibfield  {author} {\bibinfo {author} {\bibfnamefont {S.}~\bibnamefont
  {Kundu}}, \bibinfo {author} {\bibfnamefont {A.}~\bibnamefont {Hossain}},
  \bibinfo {author} {\bibfnamefont {P.~K.}\ \bibnamefont {S.}}, \bibinfo
  {author} {\bibfnamefont {R.}~\bibnamefont {Das}}, \bibinfo {author}
  {\bibfnamefont {M.}~\bibnamefont {Baenitz}}, \bibinfo {author} {\bibfnamefont
  {P.~J.}\ \bibnamefont {Baker}}, \bibinfo {author} {\bibfnamefont {J.-C.}\
  \bibnamefont {Orain}}, \bibinfo {author} {\bibfnamefont {D.~C.}\ \bibnamefont
  {Joshi}}, \bibinfo {author} {\bibfnamefont {R.}~\bibnamefont {Mathieu}},
  \bibinfo {author} {\bibfnamefont {P.}~\bibnamefont {Mahadevan}}, \bibinfo
  {author} {\bibfnamefont {S.}~\bibnamefont {Pujari}}, \bibinfo {author}
  {\bibfnamefont {S.}~\bibnamefont {Bhattacharjee}}, \bibinfo {author}
  {\bibfnamefont {A.~V.}\ \bibnamefont {Mahajan}},\ and\ \bibinfo {author}
  {\bibfnamefont {D.~D.}\ \bibnamefont {Sarma}},\ }\href
  {https://doi.org/10.1103/PhysRevLett.125.117206} {\bibfield  {journal}
  {\bibinfo  {journal} {Phys. Rev. Lett.}\ }\textbf {\bibinfo {volume} {125}},\
  \bibinfo {pages} {117206} (\bibinfo {year} {2020}{\natexlab{a}})}\BibitemShut
  {NoStop}%
\bibitem [{\citenamefont {Kittel}\ and\ \citenamefont
  {McEuen}(2005)}]{Kittel2005}%
  \BibitemOpen
  \bibfield  {author} {\bibinfo {author} {\bibfnamefont {C.}~\bibnamefont
  {Kittel}}\ and\ \bibinfo {author} {\bibfnamefont {P.}~\bibnamefont
  {McEuen}},\ }\href@noop {} {\emph {\bibinfo {title} {Introduction to solid
  state physics}}},\ \bibinfo {edition} {8th}\ ed.\ (\bibinfo  {publisher}
  {John Wiley \& Sons, Inc},\ \bibinfo {year} {2005})\BibitemShut {NoStop}%
\bibitem [{\citenamefont {Kitagawa}\ \emph {et~al.}(2018)\citenamefont
  {Kitagawa}, \citenamefont {Takayama}, \citenamefont {Matsumoto},
  \citenamefont {Kato}, \citenamefont {Takano}, \citenamefont {Kishimoto},
  \citenamefont {Bette}, \citenamefont {Dinnebier}, \citenamefont {Jackeli},\
  and\ \citenamefont {Takagi}}]{Kitagawa2018}%
  \BibitemOpen
  \bibfield  {author} {\bibinfo {author} {\bibfnamefont {K.}~\bibnamefont
  {Kitagawa}}, \bibinfo {author} {\bibfnamefont {T.}~\bibnamefont {Takayama}},
  \bibinfo {author} {\bibfnamefont {Y.}~\bibnamefont {Matsumoto}}, \bibinfo
  {author} {\bibfnamefont {A.}~\bibnamefont {Kato}}, \bibinfo {author}
  {\bibfnamefont {R.}~\bibnamefont {Takano}}, \bibinfo {author} {\bibfnamefont
  {Y.}~\bibnamefont {Kishimoto}}, \bibinfo {author} {\bibfnamefont
  {S.}~\bibnamefont {Bette}}, \bibinfo {author} {\bibfnamefont
  {R.}~\bibnamefont {Dinnebier}}, \bibinfo {author} {\bibfnamefont
  {G.}~\bibnamefont {Jackeli}},\ and\ \bibinfo {author} {\bibfnamefont
  {H.}~\bibnamefont {Takagi}},\ }\href {https://doi.org/10.1038/nature25482}
  {\bibfield  {journal} {\bibinfo  {journal} {Nature}\ }\textbf {\bibinfo
  {volume} {554}},\ \bibinfo {pages} {341} (\bibinfo {year}
  {2018})}\BibitemShut {NoStop}%
\bibitem [{\citenamefont {Li}\ \emph {et~al.}(2016)\citenamefont {Li},
  \citenamefont {Adroja}, \citenamefont {Biswas}, \citenamefont {Baker},
  \citenamefont {Zhang}, \citenamefont {Liu}, \citenamefont {Tsirlin},
  \citenamefont {Gegenwart},\ and\ \citenamefont {Zhang}}]{YMGO}%
  \BibitemOpen
  \bibfield  {author} {\bibinfo {author} {\bibfnamefont {Y.}~\bibnamefont
  {Li}}, \bibinfo {author} {\bibfnamefont {D.}~\bibnamefont {Adroja}}, \bibinfo
  {author} {\bibfnamefont {P.~K.}\ \bibnamefont {Biswas}}, \bibinfo {author}
  {\bibfnamefont {P.~J.}\ \bibnamefont {Baker}}, \bibinfo {author}
  {\bibfnamefont {Q.}~\bibnamefont {Zhang}}, \bibinfo {author} {\bibfnamefont
  {J.}~\bibnamefont {Liu}}, \bibinfo {author} {\bibfnamefont {A.~A.}\
  \bibnamefont {Tsirlin}}, \bibinfo {author} {\bibfnamefont {P.}~\bibnamefont
  {Gegenwart}},\ and\ \bibinfo {author} {\bibfnamefont {Q.}~\bibnamefont
  {Zhang}},\ }\href {https://doi.org/10.1103/PhysRevLett.117.097201} {\bibfield
   {journal} {\bibinfo  {journal} {Phys. Rev. Lett.}\ }\textbf {\bibinfo
  {volume} {117}},\ \bibinfo {pages} {097201} (\bibinfo {year}
  {2016})}\BibitemShut {NoStop}%
\bibitem [{\citenamefont {Kundu}\ \emph
  {et~al.}(2020{\natexlab{b}})\citenamefont {Kundu}, \citenamefont {Shahee},
  \citenamefont {Chakraborty}, \citenamefont {Ranjith}, \citenamefont {Koo},
  \citenamefont {Sichelschmidt}, \citenamefont {Telling}, \citenamefont
  {Biswas}, \citenamefont {Baenitz}, \citenamefont {Dasgupta}, \citenamefont
  {Pujari},\ and\ \citenamefont {Mahajan}}]{SKundu2020}%
  \BibitemOpen
  \bibfield  {author} {\bibinfo {author} {\bibfnamefont {S.}~\bibnamefont
  {Kundu}}, \bibinfo {author} {\bibfnamefont {A.}~\bibnamefont {Shahee}},
  \bibinfo {author} {\bibfnamefont {A.}~\bibnamefont {Chakraborty}}, \bibinfo
  {author} {\bibfnamefont {K.~M.}\ \bibnamefont {Ranjith}}, \bibinfo {author}
  {\bibfnamefont {B.}~\bibnamefont {Koo}}, \bibinfo {author} {\bibfnamefont
  {J.}~\bibnamefont {Sichelschmidt}}, \bibinfo {author} {\bibfnamefont
  {M.~T.~F.}\ \bibnamefont {Telling}}, \bibinfo {author} {\bibfnamefont
  {P.~K.}\ \bibnamefont {Biswas}}, \bibinfo {author} {\bibfnamefont
  {M.}~\bibnamefont {Baenitz}}, \bibinfo {author} {\bibfnamefont
  {I.}~\bibnamefont {Dasgupta}}, \bibinfo {author} {\bibfnamefont
  {S.}~\bibnamefont {Pujari}},\ and\ \bibinfo {author} {\bibfnamefont {A.~V.}\
  \bibnamefont {Mahajan}},\ }\href
  {https://doi.org/10.1103/PhysRevLett.125.267202} {\bibfield  {journal}
  {\bibinfo  {journal} {Phys. Rev. Lett.}\ }\textbf {\bibinfo {volume} {125}},\
  \bibinfo {pages} {267202} (\bibinfo {year} {2020}{\natexlab{b}})}\BibitemShut
  {NoStop}%
\end{thebibliography}%
\end{document}


\title{Supplementary Material for a novel Gapless Quantum Spin Liquid in the $4d^{4}$-honeycomb material Cu$_{3}$LiRu$_{2}$O$_{6}$ }

\author{Sanjay Bachhar}
\thanks{\href{mailto:sanjayphysics95@gmail.com}{sanjayphysics95@gmail.com}}
\affiliation{Department of Physics, Indian Institute of Technology Bombay, Powai, Mumbai 400076, India}

\author{Nashra Pistawala} 
\affiliation{Department of Physics, Indian Institute of Science Education and Research, Pune, Maharashtra-411008, India}

\author{S. Kundu}
\affiliation{Department of Physics, Indian Institute of Technology Madras, Chennai 600036, India}

\author{Maneesha Barik}
\affiliation{Department of Physics, Indian Institute of Technology Madras, Chennai 600036, India}

\author{M. Baenitz}
\affiliation{Max Planck Institute for Chemical Physics of Solids, 01187 Dresden, Germany}

\author{Jörg Sichelschmidt }
\affiliation{Max Planck Institute for Chemical Physics of Solids, 01187 Dresden, Germany}

\author{Koji Yokoyama }
\affiliation{ISIS Pulsed Neutron and Muon Source, STFC Rutherford Appleton Laboratory,Harwell Campus, Didcot, Oxfordshire OX110QX, United Kingdom}
\author{P. Khuntia}
\affiliation{Department of Physics, Indian Institute of Technology Madras, Chennai 600036, India}

\author{Surjeet Singh} 
\affiliation{Department of Physics, Indian Institute of Science Education and Research, Pune, Maharashtra-411008, India}

\author{A.V. Mahajan}
\thanks{\href{mailto:mahajan@phy.iitb.ac.in}{mahajan@phy.iitb.ac.in}}
\affiliation{Department of Physics, Indian Institute of Technology Bombay, Powai, Mumbai 400076, India}

\date{\today}
\maketitle

 Sample preparation of \ch{Cu3LiRu2O6} was done through solid-state reaction followed by soft-topotactic reaction. Structural characterization (X-ray diffraction and SEM) suggested a single phase pure compound. Then, we have measured magnetization, heat capacity and Nuclear magnetic resonance (NMR) to understand underlying physics of \ch{Cu3LiRu2O6}. Herein, we report supplementary information for a novel Quantum Spin Liquid in the honeycomb system \ch{Cu3LiRu2O6} with Ru$^{4+}$ ($4d^{4}$) .

\section{Sample Preparation}
The preparation of Cu$_{3}$LiRu$_{2}$O$_{6}$ is a two step process\cite{Huang2020}. First step is to prepare \ch{Li2RuO3}  via standard solid state reaction route. The second step is Cu-intercalation by topotactic metathesis reaction: 

\begin{equation}
	\ch{2 Li2RuO3 + 3 CuCl $\longrightarrow$ Cu3LiRu2O6 + 3 LiCl}
\end{equation}

A combination of \ch{CuCl} (Alfa Aesar, 99.9{\%}) and \ch{Li2RuO3}, in a molar ratio of 3:1, was loaded into an alumina crucible in a quartz ampoule for intercalation. The loading process was carried out in an argon-filled glove box with \ch{O2} and \ch{H2O} concentrations below 0.1 ppm. As pelletizing the sample outside the glove box was not feasible, the powder was pressed using a glass rod to increase its reactivity during topotactic reaction. The quartz ampoule was sealed under vacuum and then placed in a box furnace at a temperature of 400{\textdegree}C for 24 hours, with heating and cooling rates of 50{\textdegree}C/h. After the synthesis process, the resulting sample was washed using \ch{NH4OH} solution to eliminate any excess \ch{CuCl}, until the blue color of Cu$^{2+}$ ions disappeared. The sample was then washed five times with deionized water to remove any remaining \ch{LiCl}, which is highly soluble in water. The flame test indicated the presence of \ch{LiCl} in the by-product after filtration, which appeared as a Fuchsia Pink color, but there was no color observed under the flame test of the final washed by-product, indicating the absence of \ch{LiCl} in the sample. The sample was then dried overnight in a vacuum desiccator and dehydrated under vacuum for 8 hours at 50{\textdegree}C in a quartz ampoule to ensure complete dryness. The same process was used to prepare the nonmagnetic analog, \ch{Cu3LiSn2O6}, using \ch{Li2SnO3} as a precursor.

\section{X-ray diffraction and Rietveld Refinement}

To check the phase purity, crystal structure and elemental compositions of the sample, the powder X-ray diffraction (XRD) and SEM-EDX were performed on a  polycrystalline Cu$_{3}$LiRu$_{2}$O$_{6}$. The XRD data were collected at room temperature with Cu-K$_{\alpha}$ radiation ($\lambda=1.54182$ \AA) over the angular range 10{\textdegree} $\leq$ 2$\theta$ $\leq$ 90{\textdegree} with a 0.013{\textdegree}step size.

\begin{figure}
	\centering \includegraphics[width=1.0\linewidth]{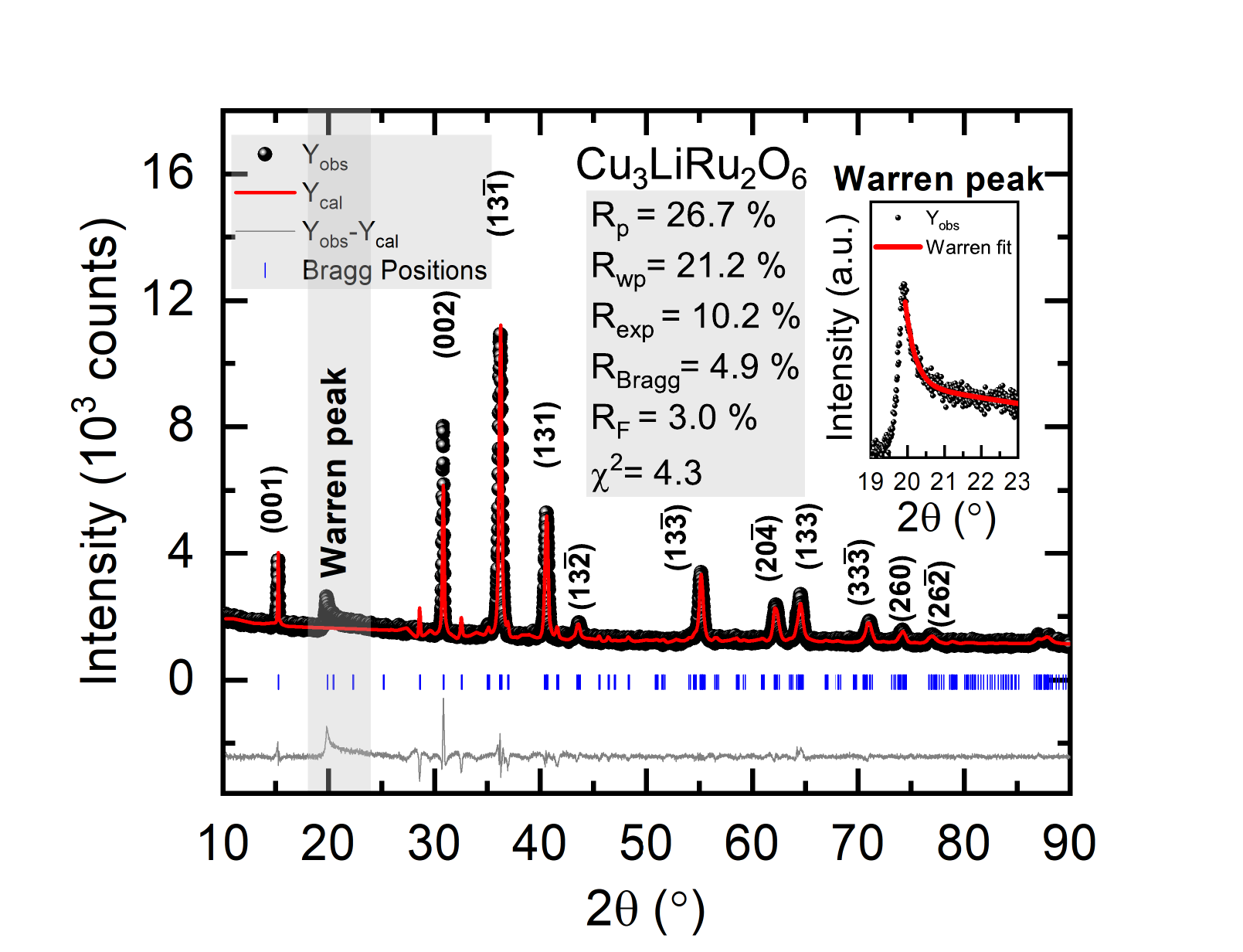}\caption{Rietveld refinement
		of powder XRD pattern of \ch{Cu3LiRu2O6} is shown along
		with Bragg positions with corresponding Miller indices (hkl). Inset
		shows the Warren peak.}
	\label{CLRO_refinement}
\end{figure}

Figure \ref{CLRO_refinement} shows the single phase Rietveld refinement of Cu$_{3}$LiRu$_{2}$O$_{6}$. From the refinement, we found that the prepared Cu$_{3}$LiRu$_{2}$O$_{6}$ crystallizes in the monoclinic C2/m space group. Table \ref{table_CLRO_1} summarizes the unit cell paramters and quality factors for the Rietveld refinement of Cu$_{3}$LiRu$_{2}$O$_{6}$. The Warren peak, a typical signature peak in 2D layered delafossite materials is observed in the range 19{\textdegree} to 23{\textdegree}, shown in inset of Figure \ref{CLRO_refinement} with a fit \cite{Bahrami_T_2019} as given below,
\begin{equation}
	I_w(2\theta)= A e^{-g(2\theta)^2}+ B/(C+(2\theta)^2),	
\end{equation}

where A,B and C are constants, $g$ is the exponent of the Gaussian term and it measures the
percentage of the stacking faults known as the $ g $-factor:
\begin{equation}
	g=\delta^2/d^2, 	\delta^2 = <d^2>-<d>^2
\end{equation}
where $ d $ is the interlayer spacing. A good fit with exponent, $ g=0.04(1) $ in the inset of Figure \ref{CLRO_refinement}  corresponds to at least 4\% volume fraction of stacking disorder. After excluding the Warren peak, Rietveld refinement is performed with faultless model (stacking faults are ignored) to extract different quality factors, atomic coordinates, site occupancy, and the isotropic Debye-Waller factors (B$_{iso}$ = 8$\pi^2$U$_{iso}$) of  Cu$_{3}$LiRu$_{2}$O$_{6}$,which are tabulated in Table \ref{table_CLRO_1} and Table \ref{table_CLRO_2} respectively.

 \begin{table}[h!]
	\caption{Unit cell parameters and quality factors are reported for the Rietveld refinement of Cu$_{3}$LiRu$_{2}$O$_{6}$ at room temperature.}
	\centering
	\scalebox{0.7}{
		\begin{tabular}{|l c| c c|}
			\hline
			Unit Cell Parameters for space group C2/m& & Quality Factors&\\
			\hline
			a(\AA) & 5.18(2) &  &  \\
			\hline
			b(\AA)& 8.92(7)& R$_{Bragg}$ (\%)&4.9 \\
			\hline
			c(\AA)& 6.05(2) & R$_F$ (\%) &3.0 \\
			\hline
			$\alpha$=$\gamma$  (\textdegree)&90 & R$_{exp}$ (\%)&10.2\\
			\hline
			$\beta$ (\textdegree)& 106.5(9) & R$_p$ (\%)&26.7\\
			\hline
			Z  & 2 & R$_{wp}$ (\%) & 21.2\\
			\hline
			V (\AA$^3$) & 268.3(5) & $\chi^2$ & 4.3\\
			\hline	
		\end{tabular}
	}
	\label{table_CLRO_1}
\end{table}

\begin{table}[h!]
	\caption{Atomic coordinates, Normalized site occupancies, and the
		isotropic Debye-Waller factors (B$_{iso}$ = 8$\pi^2$U$_{iso}$)  are reported
		for the Rietveld refinement of Cu$_{3}$LiRu$_{2}$O$_{6}$.}
	\centering
	\scalebox{0.7}{
		\begin{tabular}{|l| c| c| c| c| c| c|c|}
			\hline
			Atom & Wyckoff position & Site &x&y&z&Norm. Site Occ.&B$_{iso}$(\AA$^2$)\\
			\hline
			Ru(1)&4g& 2&0&0.333&0&2&0.4\\
			\hline
			Li(1)&2a& 2/m&0&0&0&1&0.4\\
			\hline
			O(1)&4i&m&0.416&0&0.195&2&0.5\\
			\hline
			O(2)&8j&1&0.395&0.33&0.179&4&0.5\\
			\hline
			Cu(1)&4h&2&0&0.165&1/2&1&0.5\\
			\hline
			Cu(2)&2d&2/m&1/2&0&1/2&2&0.5\\

			\hline	
		\end{tabular}
	}
	\label{table_CLRO_2}
\end{table}

\section{SEM-EDX}
 SEM-EDX stands for Scanning Electron Microscopy with Energy Dispersive X-ray spectroscopy. It is a technique that combines two analytical tools - a scanning electron microscope (SEM) and an energy-dispersive X-ray (EDX) detector - to analyze the composition of a sample. SEM-EDX is widely used in materials science, geology, and other fields to study the microstructure and elemental composition of samples. The SEM uses a beam of high-energy electrons to scan the sample's surface, creating an image with high resolution and magnification. The EDX detector then collects X-rays that are emitted from the sample's surface as a result of the interaction between the electron beam and the atoms in the sample.

\begin{figure}
\centering
\includegraphics[width=0.7\linewidth]{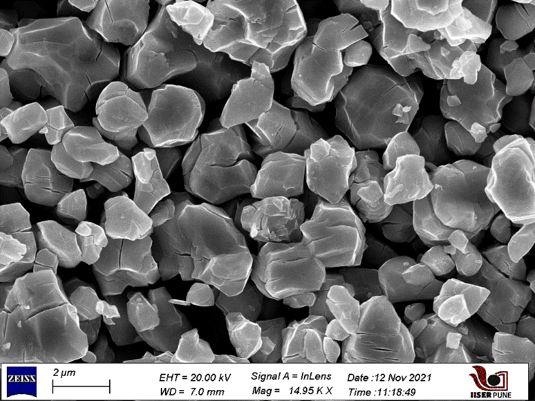}\caption{SEM image in in-lens mode at 20 kV of \ch{Cu3LiRu2O6}.}
\label{CLRO_SEM}

\end{figure}

To quantify the elemental composition, SEM-EDX measurement were carried out in in-lens mode at 20 kV on as-prepared \ch{Cu3LiRu2O6} as shown in Figure \ref{CLRO_SEM}. By analyzing the energy and intensity of the X-rays collected by the EDX detector, we found a ratio of Cu:Ru = 1.57:1 which is close to the actual composition formula of compound Cu$_{3}$LiRu$_{2}$O$_{6}$.

\section{Electron Spin Resonance}
Electron spin resonance studies were performed at X-band frequency, 9.4 GHz, using a standard continuous-wave spectrometer for temperatures between 4K and 295K. Our result could not evidence any Cu$^{2+}$ ESR line which led to the conclusion that Cu is in monovalent state.
\section{Crystal Structure}
\begin{figure}[h]
\centering{}\includegraphics[width=1.0\linewidth]{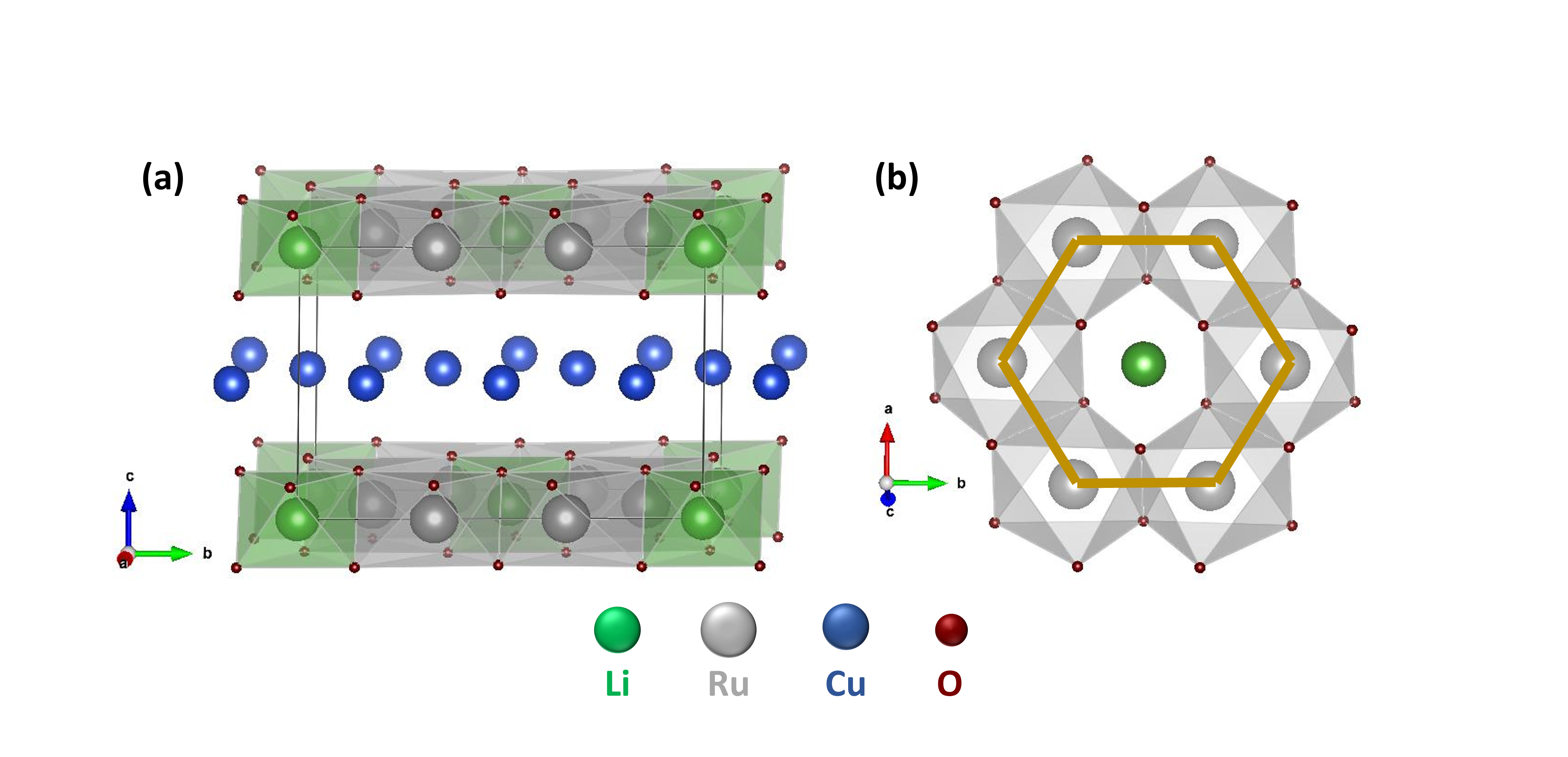}\caption{\label{fig:CLRO-schematic}{\small{}Crystal structure of Cu$_{3}$LiRu$_{2}$O$_{6}$. (a) 3D view (b) Ru-honeycomb plane}}
\end{figure}

Figure \ref{fig:CLRO-schematic}(a) shows 3D view of \ch{Cu3LiRu2O6} (through Vesta software) spacer atom Cu separates two Ru-honeycomb plane where Li placed at the center of honeycomb. The Ru-honeycomb ab-planar view with Triplon flavor (Red, blue and green colour bond like Kitaev model) is shown in Figure \ref{fig:CLRO-schematic}(b). Edge shared \ch{RuO6} octahedra leads to 90{\textdegree} $dpd$ geometry with two indirect hopping paths. Structural analysis suggest a perfect honeycomb
with 2.98 {\AA} each $Ru-Ru$ bond with $Ru-O-Ru$ bond angle 94.9{\textdegree}. Electron Spin
Resonance suggests Cu is non-magnetic with Cu$^{1+}$ and hence, source of magnetism is
Ru$^{4+}$. This geometry with Ru$^{4+}$ ($4d^{4}$) is a promising system to explore novel excitonic magnetism.

\section{Magnetization}
The dc magnetization $M(T)$ as a function of temperature $T$ in the range 2-600 K was measured on a cold pressed pellet of  Cu$_{3}$LiRu$_{2}$O$_{6}$ in ZFC mode in a 10 kOe field. Further, low field (100 Oe), ZFC/FC measurement were also carried out. $M$ vs $H$ measurments were done at 20 K to check sample quality. We carried out magnetization measurements using a Quantum Design Magnetic Property Measurement System (MPMS) and Vibrating Sample magnetometer (VSM).


\begin{figure}[h]
	\centering{}\includegraphics[width=1.0\linewidth]{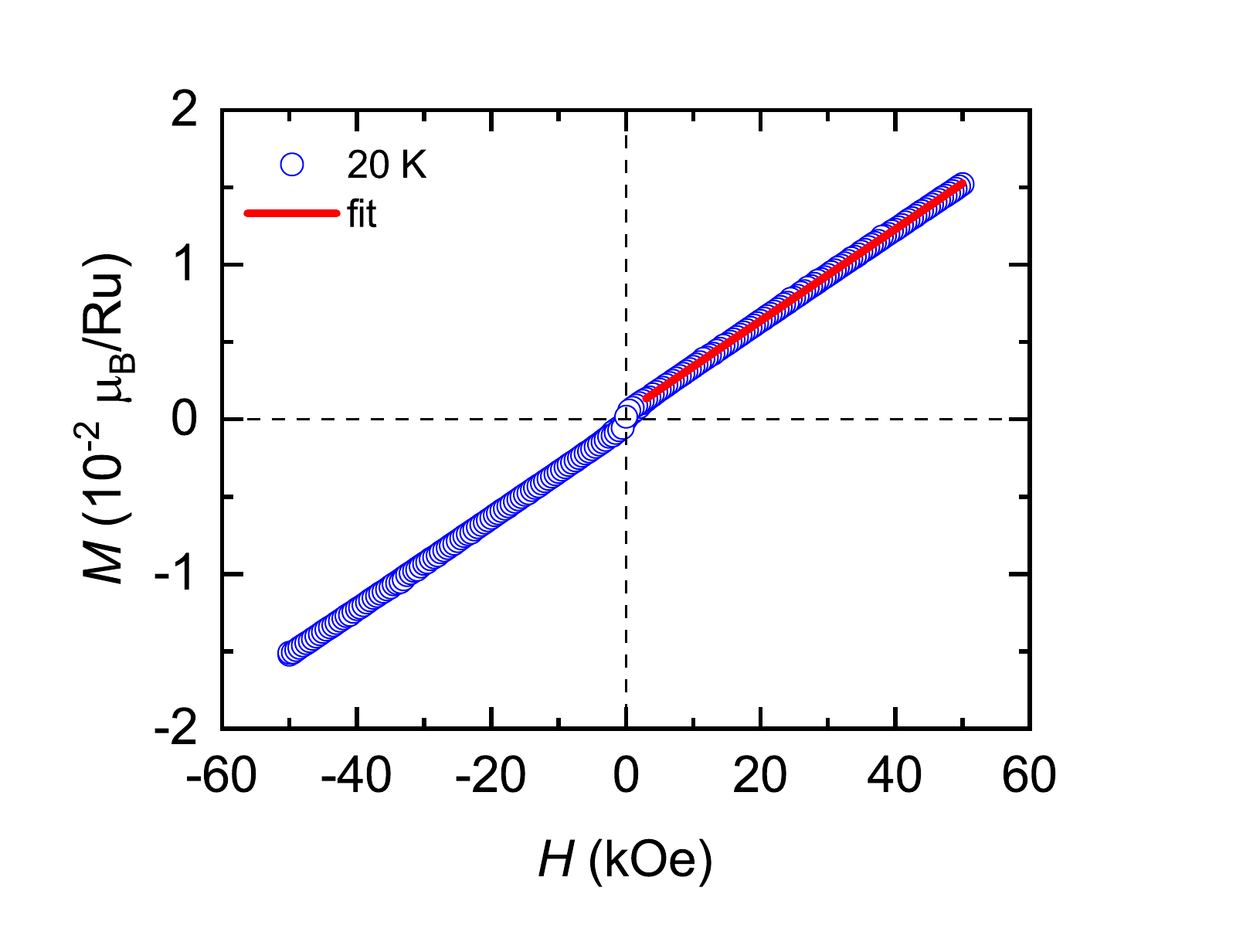}\caption{\label{CLRO_MH_20K}{\small{}Plot of $M$ vs $H$ at 20 K. The red line is a linear fit to estimate the impurity concentration.}}
\end{figure}


  For \ch{Cu3LiRu2O6}, we do not observe a perfect linear behavior magnetization in field (Figure \ref{CLRO_MH_20K}) but change in slope around origin. Saturation magnetization, M$_{sat}$ due to ferromagnetic impurity can be estimated by the formula, expressed as  
 \begin{equation}
 	M(H) = M_{sat} + \chi H	
 	\label{Msat_cal}
 \end{equation}            
 
 where M$_{sat}$ is the saturation magnetization, $\chi$ is magnetic susceptibility. Figure \ref{CLRO_MH_20K} shows fitting the data to equation \ref{Msat_cal}. Fitting yields M$_{sat} = 4.5 \times 10^{-4} \mu_{B}/Ru$. For Fe (iron), saturation magnetization is 2.2 $\mu_{B}$/Fe. We obtained impurity amount by taking ratio of M$_{sat}$ (calculated) and M$_{sat}$ (Fe). We found $\sim$ 200 ppm impurity.


\begin{figure}
	\centering
	\includegraphics[width=1.0\linewidth]{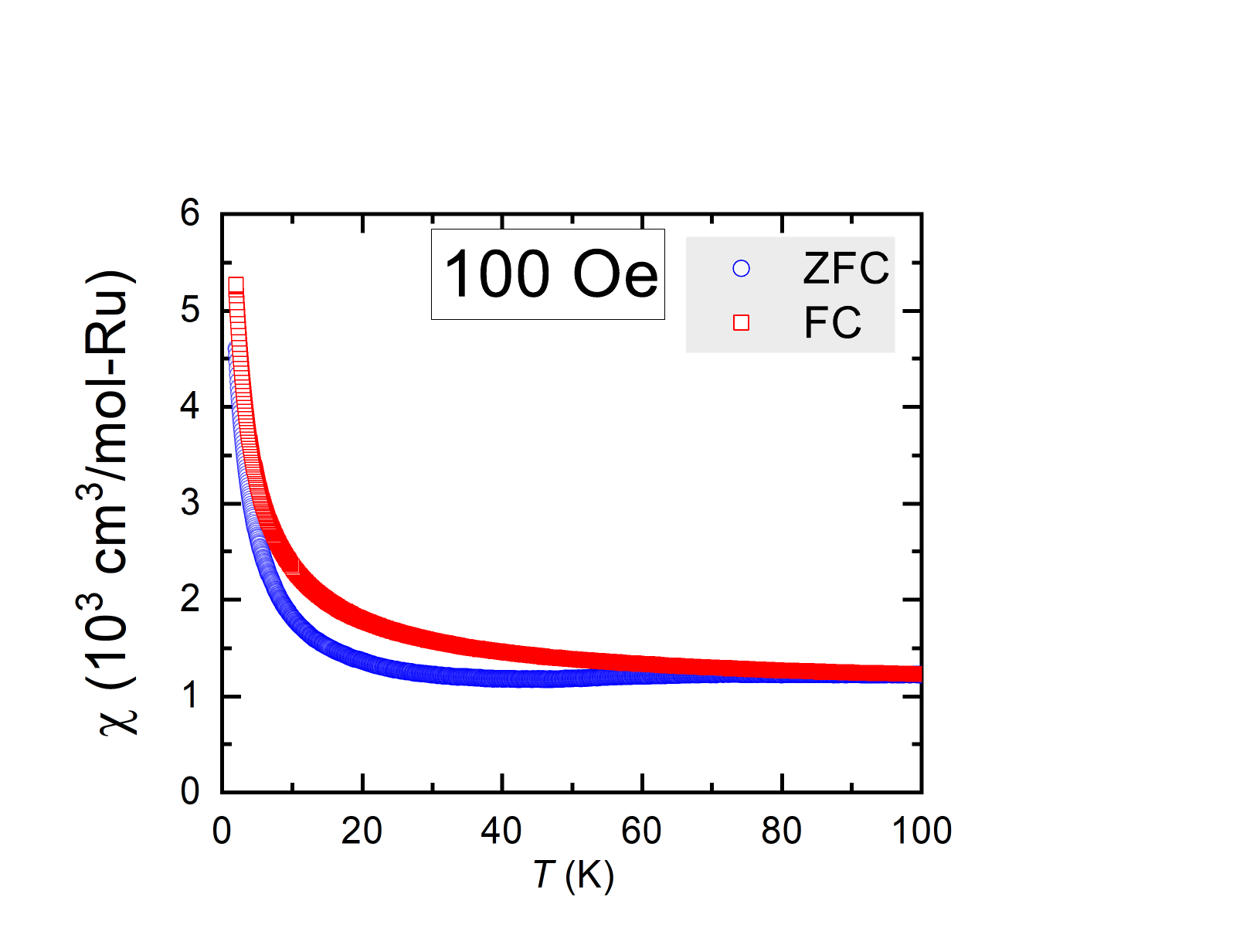}\caption{The semi-log plot of $\chi$(T) vs. T between zero field cooled (ZFC) (red triangle) and field cooled (FC) (blue circle) mode at an applied field H=100 Oe in the temperature range 100 K to 2 K. Bifurcation indicates presence of static moments (possibly extrinsic).}
	\label{CLRO_MT_100Oe}
\end{figure}

\begin{figure}
	\centering
	\includegraphics[width=1.0\linewidth]{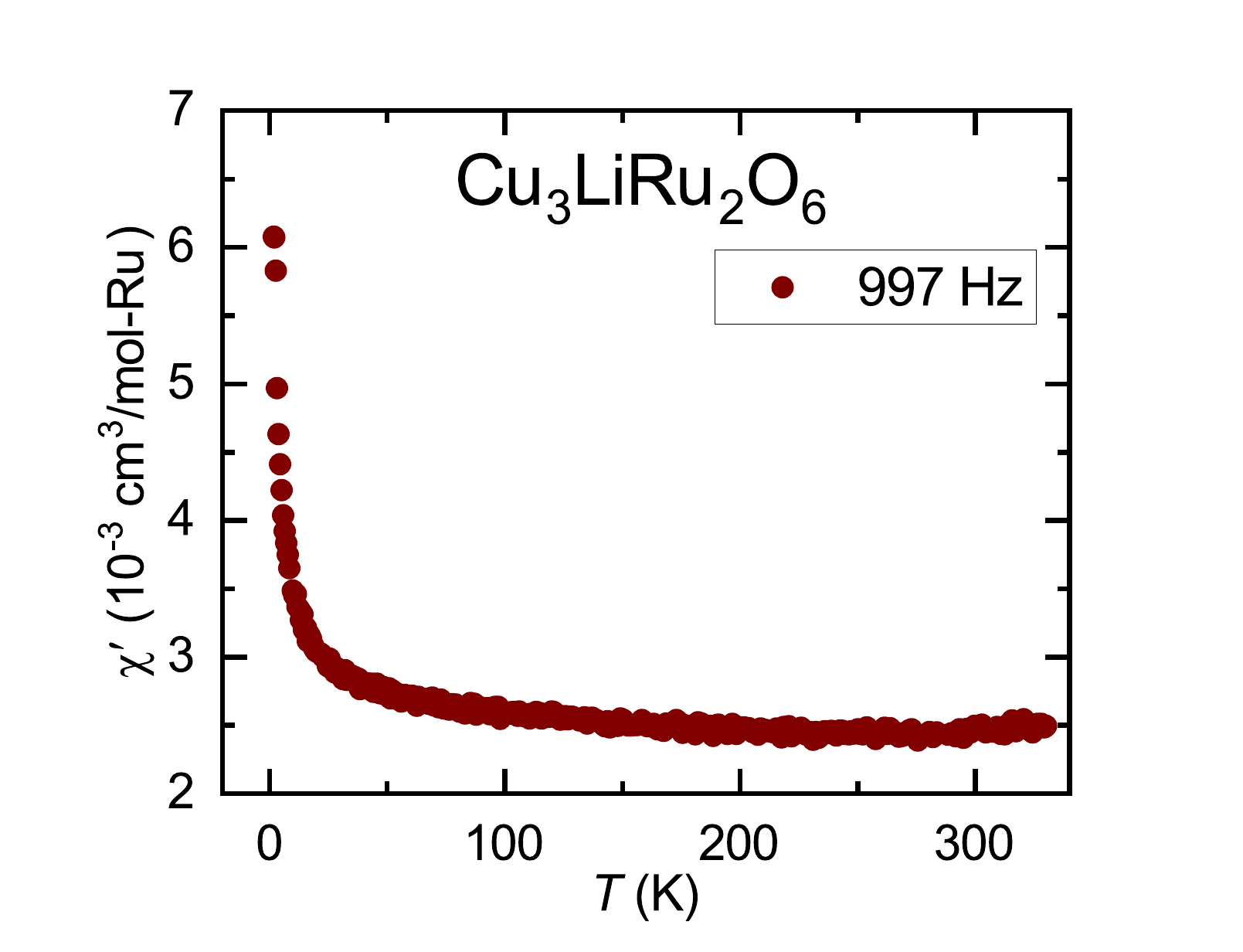}\caption{Temperature dependence of the real part of ac susceptibility ($\chi$), measured at 997 Hz with zero external dc magnetic field and an ac field of 1 Oe.}
	\label{CLRO_AC_chi}
\end{figure}


We corrected the measured magnetization by subtracting M$_{sat}$ due to impurity. Then, we calculated $\chi$ for \ch{Cu3LiRu2O6}. Bulk magnetic measurements were conducted on \ch{Cu3LiRu2O6} from 600 K to 2 K (Main paper).We observed magnetic susceptibility (blue sphere) rising from 600 K to 300 K which was of Curie-Weiss nature. Around $\sim$ 300 K, there was a broad anomaly followed by Curie-tail at low-T. The peak at 300 K could be due to short range spin correlation (signature peak) in 1D chain of honeycomb lattice. The origin of peak might be also connected to antiferromagnetic interaction strength ($\theta_{CW}= - 222 K$) in the honeycomb lattice, analogous to \ch{H3LiIr2O6}. There were no anomalies (except 300 K) down to 2 K, which ruled out the presence of any long-range ordering within the system. Fitting to the data (400-600 K) with Curie-Weiss equation yields $\chi_{0} = 1.0 \times 10^{-6}$ (cm$^{3}$/mol Ru), C = 0.73 (K cm$^{3}$/mol Ru), and $\theta_{CW} = -222$ K. The effective moment was determined as $\mu_{\text{eff}} = 2.4 \mu_{B}$, corresponding to $J_{\text{eff}}=1$ or $S=1$. Analysis of the inverse susceptibility plot indicated strong antiferromagnetic correlations among Ru$^{4+}$ ions, while the frustration parameter, $f =\frac{\mid\theta_{CW}\mid}{T_{N}} \sim 4400$, suggested that \ch{Cu3LiRu2O6} is a highly frustrated system. 



The DC susceptibility data in ZFC and FC mode at 100 Oe field plotted in semi-log scale, shown in Figure \ref{CLRO_MT_100Oe}. ZFC-FC bifurcation below 100 K indicates presence of static moments in system though extrinsic in origin. As we have not seen any glassy transition in AC-susceptibility ($\chi_{AC}$) measurement of \ch{Cu3LiRu2O6} as shown in figure \ref{CLRO_AC_chi}, we conclude that ZF-FC bifurcation is due to extrinsic impurity and absence of static moments in the system which is also supported by $\mu$SR measurements on \ch{Cu3LiRu2O6}.

\section{Heat Capacity}
To get more information about low-energy excitations, we measured the heat capacity of the sample at constant pressure C$_{p}$(T) at different fields (0-90 kOe) with temperature range (400 mK-300 K) for 0 kOe and (400 mK-40 K) for other fields.  

In general total specific heat, C$_{p}$ of the system can be expressed as:

\begin{equation}
	C_{p}(T,H)= C_{lattice}(T) + C_{Sch}(T,H) + C_{m}(T,H)
\end{equation}

The lattice specific heat is denoted as $C_{lattice}$, while $C_{Sch}$ represents the Schottky contribution caused by independent paramagnetic spins present in the system. The remaining magnetic contribution to the specific heat is referred to as $C_{m}$. Our main objective is to determine the "intrinsic" magnetic contribution $C_{m}$, which is independent of C$_{lattice}$ and C$_{Sch}$. To obtain the intrinsic $C_{m}$, we must first calculate C$_{lattice}$ and C$_{Sch}$, and then deduct them from the total specific heat $C_{p}$. There are a few ways to calculate $C_{lattice}$. A prominent way is to measure heat capacity of a suitable non-magnetic analog and another is Debye-Einstein fitting\cite{SKundu2020_YCTO}.

\begin{figure}
	\centering
	\includegraphics[width=0.9\linewidth]{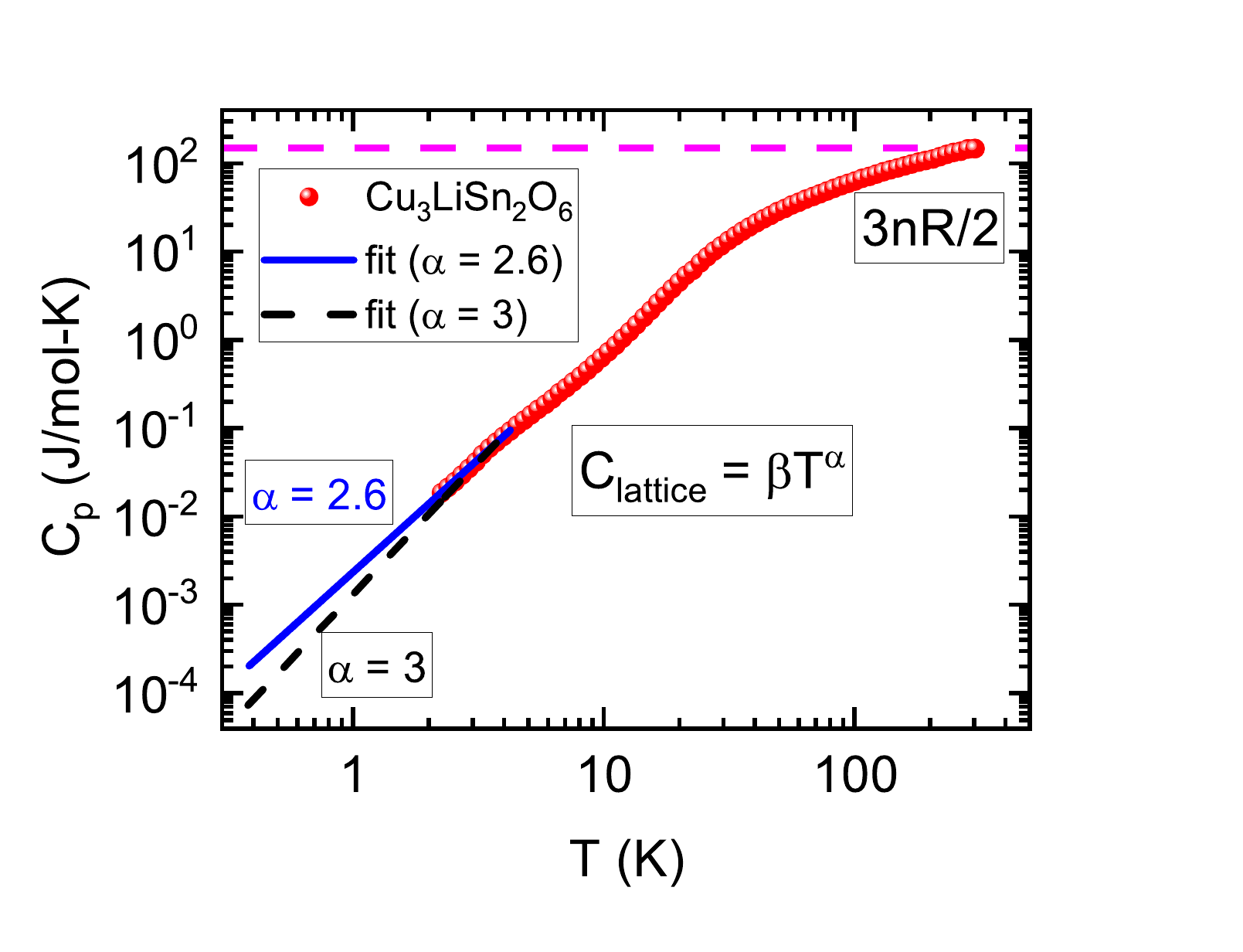}
	\caption{Specific heat, C$_{p}$ per Sn as a function of temperature for \ch{Cu3LiSn2O6} (red sphere). Blue solid line and black dot line are fitted curve with $\alpha=2.6$ and $\alpha=3$ respectively.}
	\label{CLSO_fit}
\end{figure}

\begin{figure}
	\centering
	\includegraphics[width=0.9\linewidth]{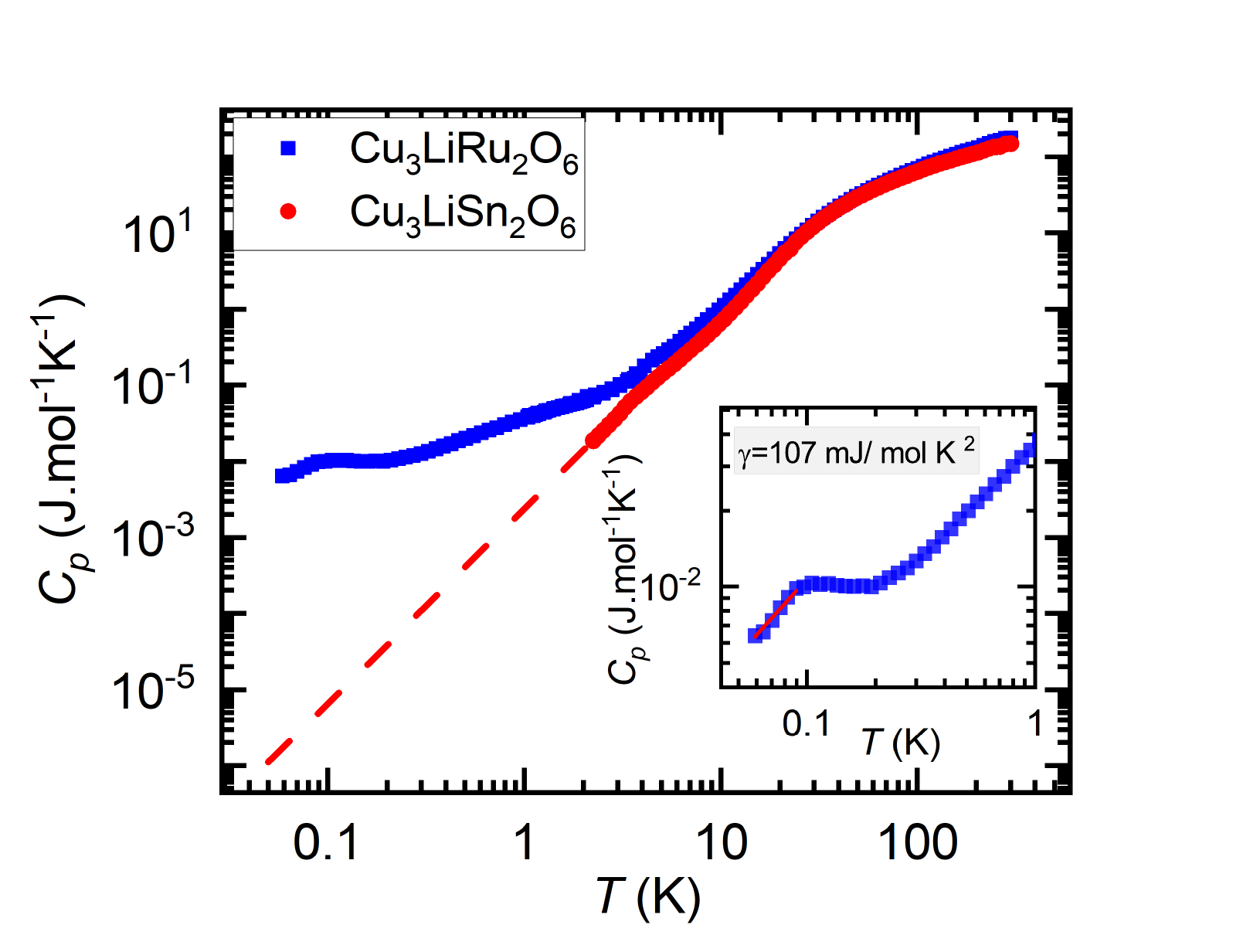}
	\caption{Specific heat, C$_{p}$ per Ru as a function of temperature for \ch{Cu3LiRu2O6} (blue square) and its non-magnetic analog \ch{Cu3LiSn2O6} (red circle). (Inset) Cp (60 mK to 1 K range) has shown with linear ($\sim T$) fit below 100 mK.}
	\label{Cp_CLRO}
\end{figure}

 We have a suitable non-magnetic analog \ch{Cu3LiSn2O6} (see figure \ref{CLSO_fit}) as a heat capacity reference of \ch{Cu3LiRu2O6}. We have measured the heat capacity of \ch{Cu3LiSn2O6} from 300 K to 2 K. At high temperature, value of C$_{p}$ should be 3nR (n is number of atoms per fomula unit, R is gas constant) according to Dulong-Petit law\cite{Kittel2005}. We plotted $C_{p}$ per Sn in figure \ref{CLSO_fit}, so we expected 3nR/2 at 300 K. We observe $C_{p}$ of \ch{Cu3LiSn2O6} at 300 K matching with 3nR/2 reference line (pink dash line).  At low-T, lattice heat capacity should follow power law ($C_{lattice}\sim \beta T^{3}$) theoretically (black dotted line is shown in figure \ref{CLSO_fit}). We fitted data to $C_{lattice} \sim \beta T^{\alpha}$ in the temperature range 2-4 K and it yields $\beta=0.002$ and $\alpha=2.6$. Then, we extrapolated the data from 2 K to 400 mK for $C_{m}$ calculation. Note that at 400 mK $C_{p}$ of \ch{Cu3LiSn2O6} value $\sim$ 10$^{-4}$ which is 1/100 th of measured $C_{p}$ of \ch{Cu3LiRu2O6} (that is $\sim$ 10$^{-2}$). So, small error ($\Delta$ $\alpha$ $=$ 0.4) in $C_{lattice}$ at low-T does not affect to calculate $C_{m}$.

Figure \ref{Cp_CLRO} shows $C_{p}$ with $T$ down to 60 mK for \ch{Cu3LiRu2O6}, together with its non-magnetic analog \ch{Cu3LiSn2O6}. No sign of long range ordering down to 60 mK. Linear $T$-dependence at low-$T$ indicates gapless fermionic-like excitations. We fitted the data with $C_{p} = \gamma T$ (Inset of Figure \ref{Cp_CLRO}) and obtained Sommerfeld coefficient, $\gamma=107$ mJ/mol K$^{2}$ which is much larger than that in Fermi liquid.

\begin{figure}
	\centering
	\includegraphics[width=1.0\linewidth]{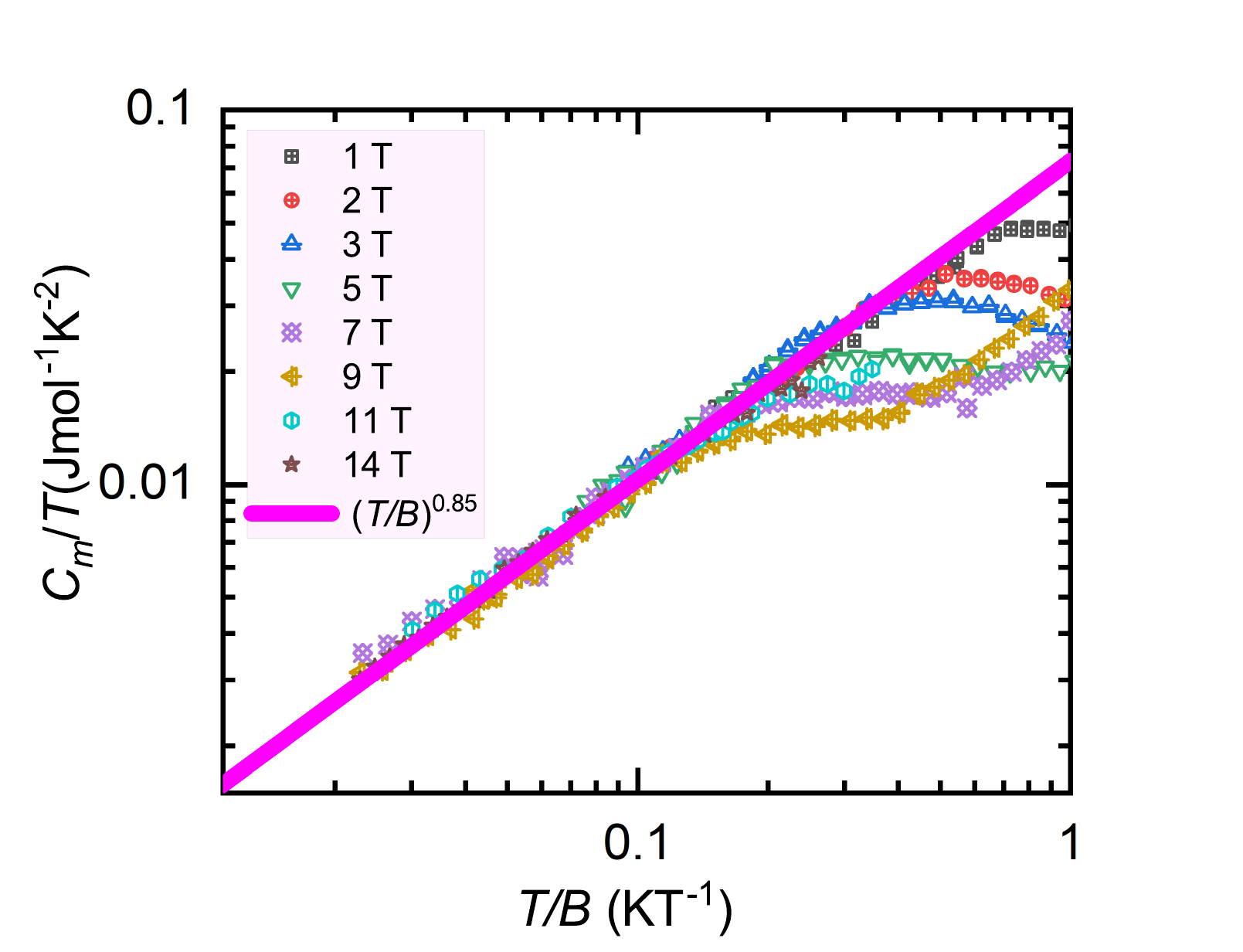}
	\caption{The data collapsed/scaling of $C_{m}$: $C_{m}/T$ vs $T/B$. The pink line corresponds to $(T/B)^{0.85}$ line.}
	\label{Cm_scaling}
\end{figure}

 Magnetic heat capacity, C$_{m}$ as a function of temperature for Cu$_{3}$LiRu$_{2}$O$_{6}$ in various fields is shown in main paper. A power law variation of C$_{m}$ at low-$T$ indicates gapless excitations which is consistent with other experiments. On application of a field, the exponent changes from 1 to 1.8 which might be due to presence of vacancy induced states in the Quantum Spin Liquid. At low-$T$, $C_{m}$ is dependent on the applied field and scaling of $C_{m}$ data (Figure \ref{Cm_scaling}) is observed which could be due to the effect of the field on the defect-induced low-energy density-of-states. We speculate that the scaling originates from small amount of vacancy presence in the system and leads to possible local random singlet phase on top of Quantum Spin Liquid.

\begin{figure}
 	\centering
 	\includegraphics[width=1.0\linewidth]{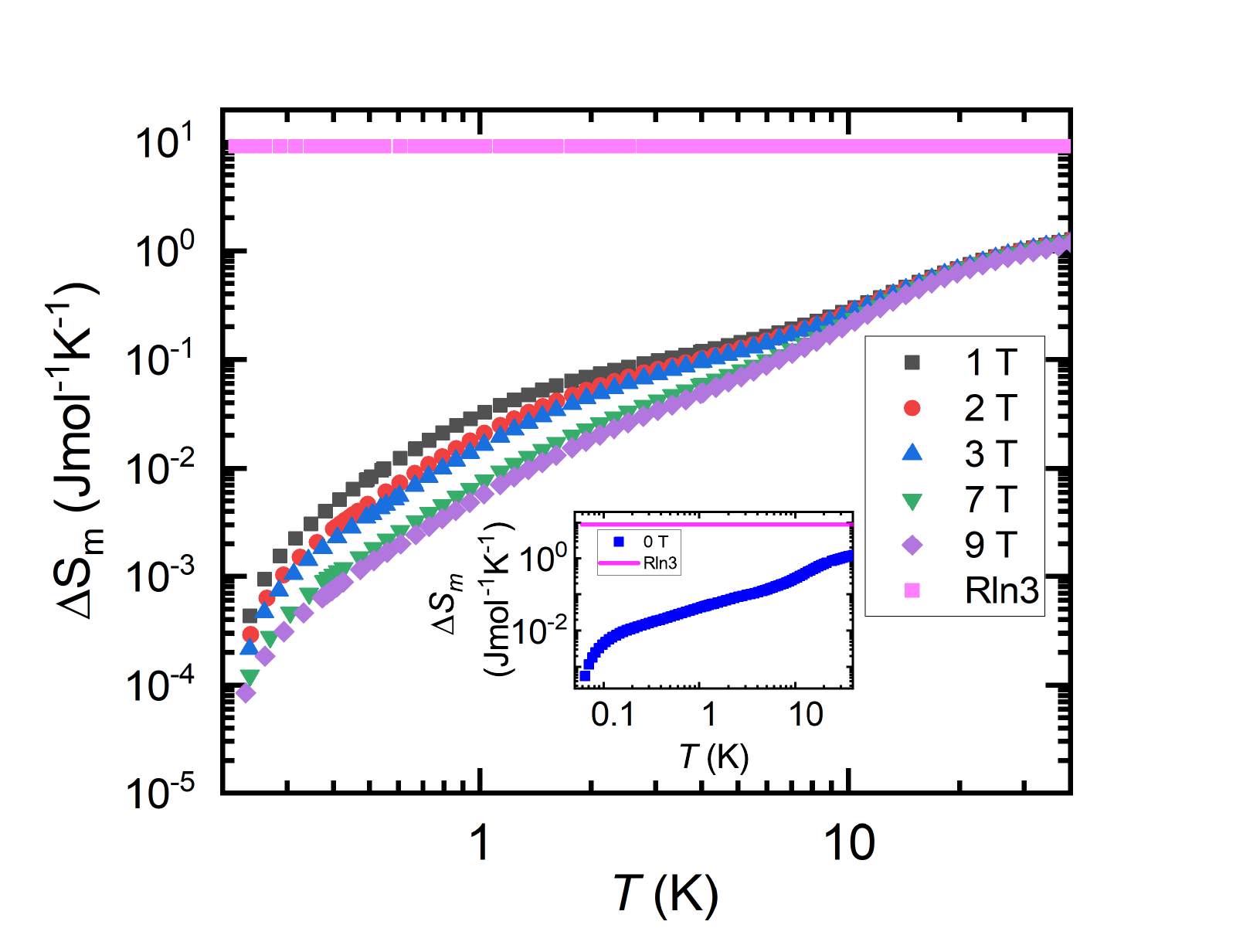}
 	\caption{Magnetic entropy, $\Delta$S$_{m}$ as a function of temperature at various fields.}
 	\label{Sm_CLRO}
 \end{figure}

 After obtaining $C_{m}$, we proceeded to compute the magnetic entropy change $\Delta S_{m}$ (given by $\Delta S_{m} = \int \frac{C_{m}}{T}dT$) for the system under various applied fields: 0 kOe, 10 kOe, 40 kOe and 90 kOe (Figure \ref{Sm_CLRO}). For a $S=1$ system, the theoretical maximum entropy change is R$\ln$(2J$_{eff}$ + 1) = 9.13 J mole$^{-1}$ K$^{-1}$ (Pink line in Figure \ref{Sm_CLRO}). By using the formula ($\frac{\Delta S_{m}}{Rln3} \times 100$), we calculated maximum entropy recovery upto 40 K. The results showed a maximum entropy recovery is approximately 14{\%} of Rln3 for S=1 system, indicating that the system contains numerous low-lying degenerate ground states and is therefore highly frustrated which is consistent with frustration parameter ($f \sim 4400$) inferred from bulk-$\chi$. 
 
\section{Nuclear Magnetic Resonance}

The nucleus $^{7}$Li is an ideal candidate for NMR investigations due to its I = 3/2 nuclear spin and 92.6{\%} natural abundance. The gyromagnetic ratio of $^{7}$Li is 16.54 MHz/Tesla. For our NMR study of the \ch{Cu3LiRu2O6} sample, we utilized a Cu-coil with a frequency range of 90-180 MHz, and the sample's mass was 282 mg. We exposed the sample to a homogeneous magnetic field of approximately 93.95 kOe and excited the target nucleus with a radio frequency (rf) pulse equal to its Larmor frequency ($\omega_{0}$). The spectra were obtained by Fourier transforming the spin-echo after a $\pi/2-\pi$ pulse sequence. We determined the spin-lattice relaxation time (T$_{1}$) using a saturation recovery pulse sequence with a $\pi/2$ pulse of 3 $\mu$s.

\subsection{$^{7}$Li NMR Spectra}

In order to probe the intrinsic spin susceptibility behavior of the correlated Ru$^{4+}$ ions on honeycomb lattice in Cu$_{3}$LiRu$_{2}$O$_{6}$, the $^{7}$Li (I=3/2, $\gamma$/2$\pi$ = 16.546 MHz/T) NMR spectra measurements were performed at a fixed magnetic field (93.954 kOe) in the temperature range: 300 K to 80 K at IIT Bombay and further low temperature measurements were continued at 90 kOe field sweep magnet in Max Planck Institute for Chemical Physics of Solids in the temperature range: 110 K to 2 K.  

\begin{figure}
	\centering
	\includegraphics[width=1.2\linewidth]{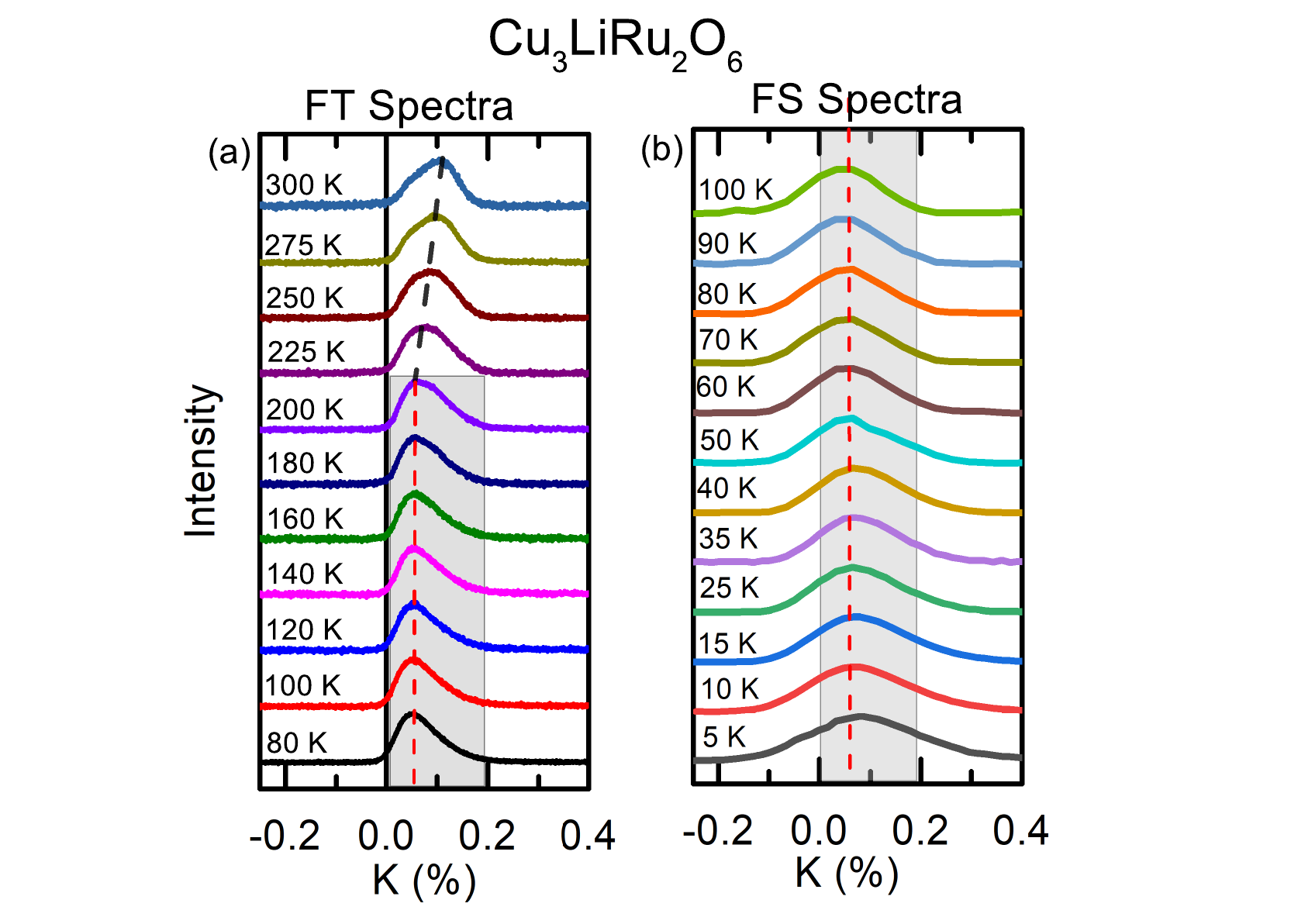}
	\caption{$^{7}$Li NMR Spectra [(a) Fourier Transform Spectra and (b) Field Sweep Spectra] for \ch{Cu3LiRu2O6} as a function of Knight shift. Dotted line is the guide to the peak positions of peak.}
	\label{Spectra_CLRO_Sweep_FT}
\end{figure}
Figure \ref{Spectra_CLRO_Sweep_FT}(a) shows $^7$Li NMR Spectra for Cu$_{3}$LiRu$_{2}$O$_{6}$ as a function of NMR shift (reference position was at 155.45741 MHz, that of non-magnetic analog: Cu$_{3}$LiSn$_{2}$O$_{6}$) in the temperature range (80-300 K). The black dotted line is the guide to eye for peak position of the spectra in the temperature range 300 K to 200 K. The peak position shifts towards the reference line (a solid black line at K=0) with decreasing temperature as shown in Figure \ref{Spectra_CLRO_Sweep_FT}. The red dotted line is guide to eye for peak positions of spectra in the temperature range 200 K to 80 K. The peak position remains unchanged with same spectral line width in this temperature range. Figure \ref{Spectra_CLRO_Sweep_FT}(b) shows field Swept spectra as function of NMR shift in the temperature range 100 K to 5 K. We have remeasured 80-100 K spectra for comparison with our fixed field Fourier transform spectra. NMR shift data in a persistent magnet is more accurate than Field sweep magnet so we carried forward NMR shift of fixed magnetic field (93.954 kOe) down to 2 K by overlapping common temperature spectra (80 K and 100 K). The red dotted line is a guide to eye for peak positions of spectra in the temperature range 100 K to 5 K. The peak position continues to remain the same from 100 K to 5 K. The shape of spectra is  asymmetric in nature probably due to non-uniform internal field distribution.

The NMR spectral shift is proportional to intrinsic spin susceptibility, $\chi_{spin}$ and hyperfine coupling, $A_{hf}$. The temperature dependent variation of $^{7}$K for \ch{Cu3LiRu2O6} is similar to that observed in \ch{H3LiIr2O6} except shift in high temperature peak position (See figure in main paper). The reason of shift in peak from 140 K (for \ch{H3LiIr2O6}) to 300 K (for \ch{Cu3LiRu2O6}) could be due to change in exchange energy. $^{7}$K has downfall below 300 K to 200 K and then becomes flat from 200 K to 2 K with $\sim$ 0.05{\%}, a significant non-zero value. The temperature independent $^{7}$K variation below 200 K indicates intrinsic susceptibility is temperature independent and bulk susceptibility below 200 K is strongly dominated by impurity or free spins. The reason of 140 K peak in $^{7}$K vs T for \ch{H3LiIr2O6} are not quite clear from Literature\cite{Kitagawa2018}. However, the peak is not associated to magnetic ordering but possibly connected to antiferromagnetic interaction strength ($\theta_{CW}= - 105$ K) and honeycomb lattice. In \ch{Cu3LiRu2O6}, there is 300 K broad anomaly in bulk susceptibility  as well as $^{7}$K. The origin of peak could be connected to antiferromagnetic interaction strength ($\theta_{CW}= - 222$ K) and honeycomb lattice. 

\subsection{Spin-lattice relaxation}
Due to the minor electric quadrupole splitting of $^{7}$Li, the NMR relaxation rate T$_{1}$ was determined using the saturation recovery technique  and fitting the data to an empirical stretched-exponential recovery function of the form:

\begin{equation}\label{saturation}
	\begin{split}
		1- (M(t_d)/ M(0)) & =  A* exp(-(t_d/T_1)^{\beta}) \\ 
		&   \\
	\end{split}
\end{equation}

here T$_{1}$ is the spin-lattice relaxation, $\beta$ is the stretching exponent, A is the saturation level, M(t$_{d}$) is the magnetization at delay time t$_{d}$, M(0) is the saturation magnetization.

\begin{figure}
	\centering
	\includegraphics[width=1.0\linewidth]{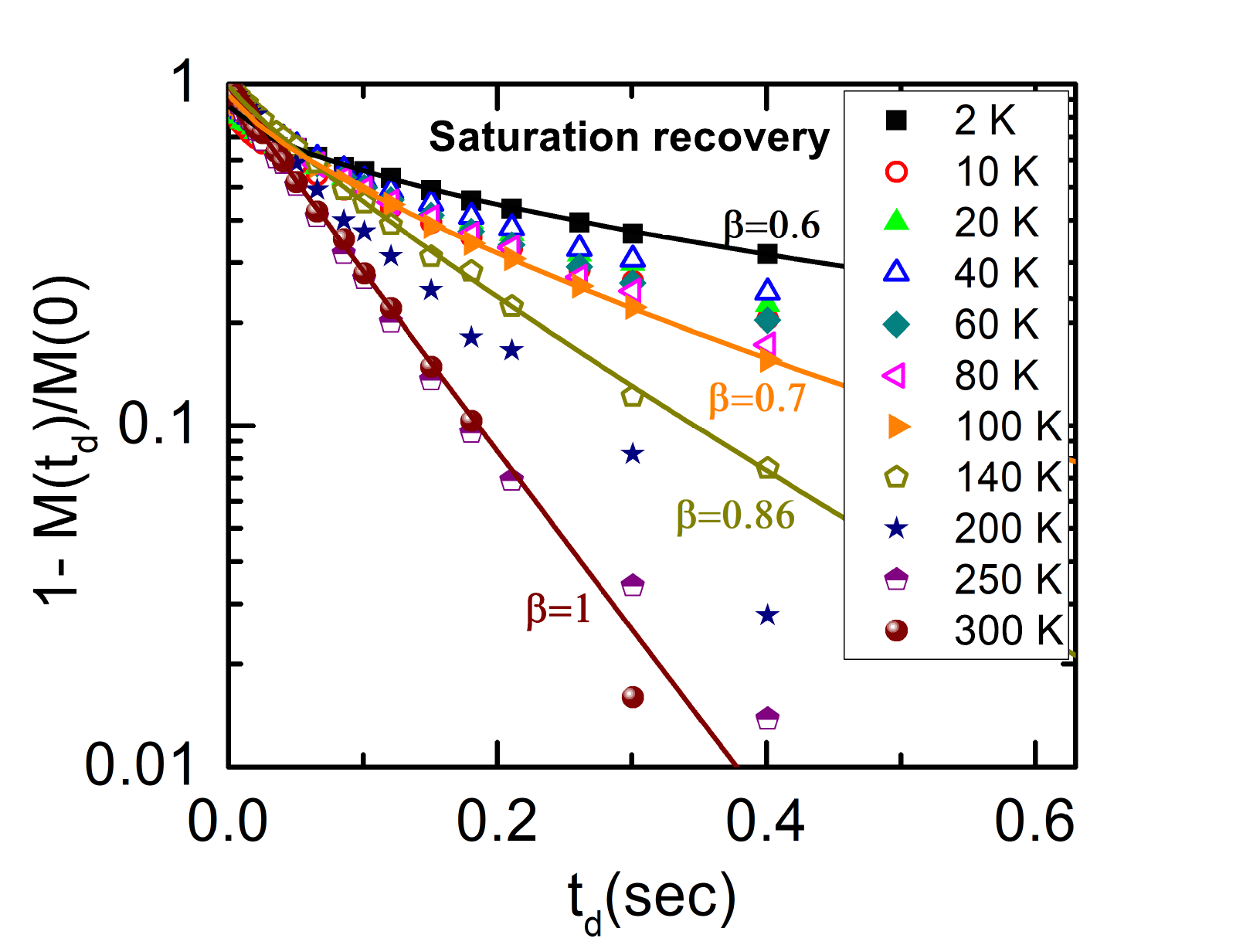}
	\caption{Recovery of the longitudinal nuclear magnetization as a function of delay time ($t_{d}$) between the saturation $\pi$/2 pulse and the probe ($\pi$/2 - $\pi$) pulses.}
	\label{CLRO_normM}
\end{figure}

The Figure \ref{CLRO_normM} shows a few selected plots of $1- (M(t_d)/ M(0))$ vs $t_{d}$ in the temperature range 2-300 K to demonstrate the T$_{1}$ calculation process. Above 250 K, the saturation recovery can be fitted with single exponential with $\beta=1$. Below 250 K, $\beta$ decreases and becomes flat below 60 K, as shown in main paper. This deviation of $\beta$ from the ideal value ($=1$) can be attributed to the distribution of internal states in space, possibly caused by the impact of magnetic defects that are distributed randomly. In frustrated magnets such as quantum spin liquids, a value of $\beta \sim 0.5$ is commonly observed, which corresponds to randomized dipolar hyperfine fields. The spin lattice relaxation rate, $1/T_{1}$ as a function of temperature. $1/T_{1}$ measures the $q$-averaged imaginary part of the dynamical spin susceptibility, $\chi"(q,\omega_{0})$. Dynamical spin susceptibility is a measure of the response of a material's magnetic moment to an applied magnetic field. Specifically, it is the Fourier transform of the autocorrelation function of the spin fluctuations in the material. Power law variation ($\sim T^{1}$) of spin-lattice relaxation at low-T indicates gapless excitation, often seen in Quantum Spin Liquids.

\section{Muon Spin Relaxation($\mu$SR)}

We measured muon asymmetry at zero-field (ZF) in the temperature range 50 K and down to 50 mK at the ISIS, UK facility. No oscillations in ZF-muon asymmetry down to 50 mK. That indicates  there is no long range magnetic ordering down to 50 mK.  We could fit the muon relaxation asymmetry well using $ A(t)= A_{rel} G_{KT}(\Delta,T) exp(- \lambda t) + A_{0} $. Here, $G_{KT}(\Delta,T)$ is the Kubo-Toyabe function which models the relaxation of muons in a Gaussian distribution of magnetic fields from nuclear moments and $A_{rel}$ is the relaxing asymmetry. From these fits we obtain the static nuclear field distribution ($\Delta$) to be about 1.8 Oe. This value is typical of nuclear dipolar fields at the muon site, in the present case arising from $^{63,65}$Cu, $^{6,7}$Li and $^{99,101}$Ru nuclei. The exponential term $exp(- \lambda t)$ arises from the relaxation due to fluctuations of the electronic local moments. Below 3 K, we well fitted data with a stretch exponential in addition to a constant i.e., $A(t)=A_{rel} exp(-(\lambda t)^\beta) + A_{0}$. Here, $\beta$ is stretching exponent which often seen in non-uniform magnetism. Below about 3 K, $\beta$ deviates from 1 and becomes constant with value $\sim$ 0.7 below 1 K and down to 0.05 K. Below 1 K, muon spin relaxation rate, $\lambda$ saturates down to 0.05 K. That indicates persistent dynamics presence in the system. Next, We measured longitudinal field (LF) measurement at various fields to check how the muon decouples from the internal fields.

\begin{figure}
	\centering
	\includegraphics[width=1.0\linewidth]{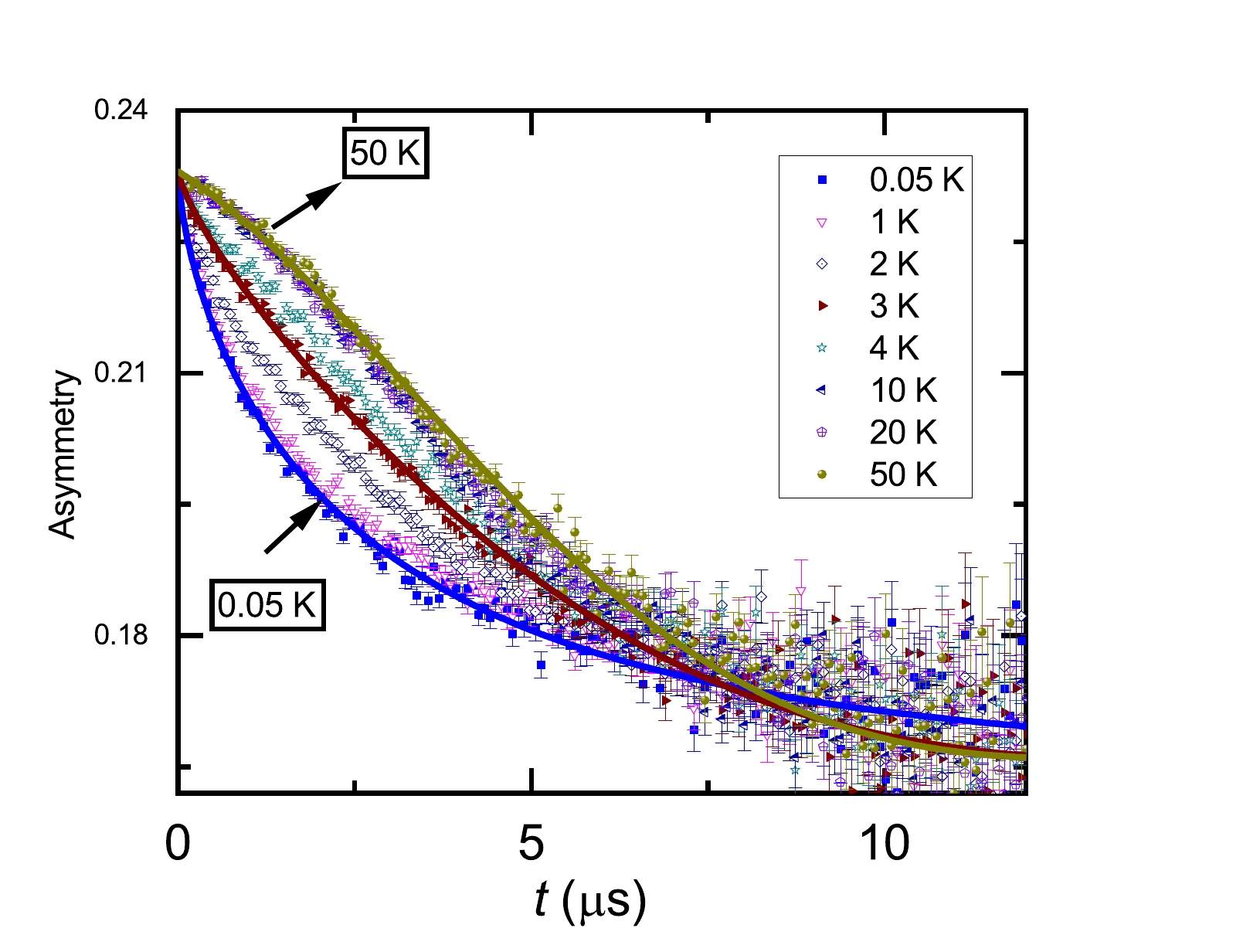}
	\caption{The variation of zero field muon asymmetry with time at various temperatures (50 K to 50 mK) for \ch{Cu3LiRu2O6}. The solid lines are fit as described in the text.}
	\label{ZF_asym_CLRO}
\end{figure}

\begin{figure}
	\centering
	\includegraphics[width=1.0\linewidth]{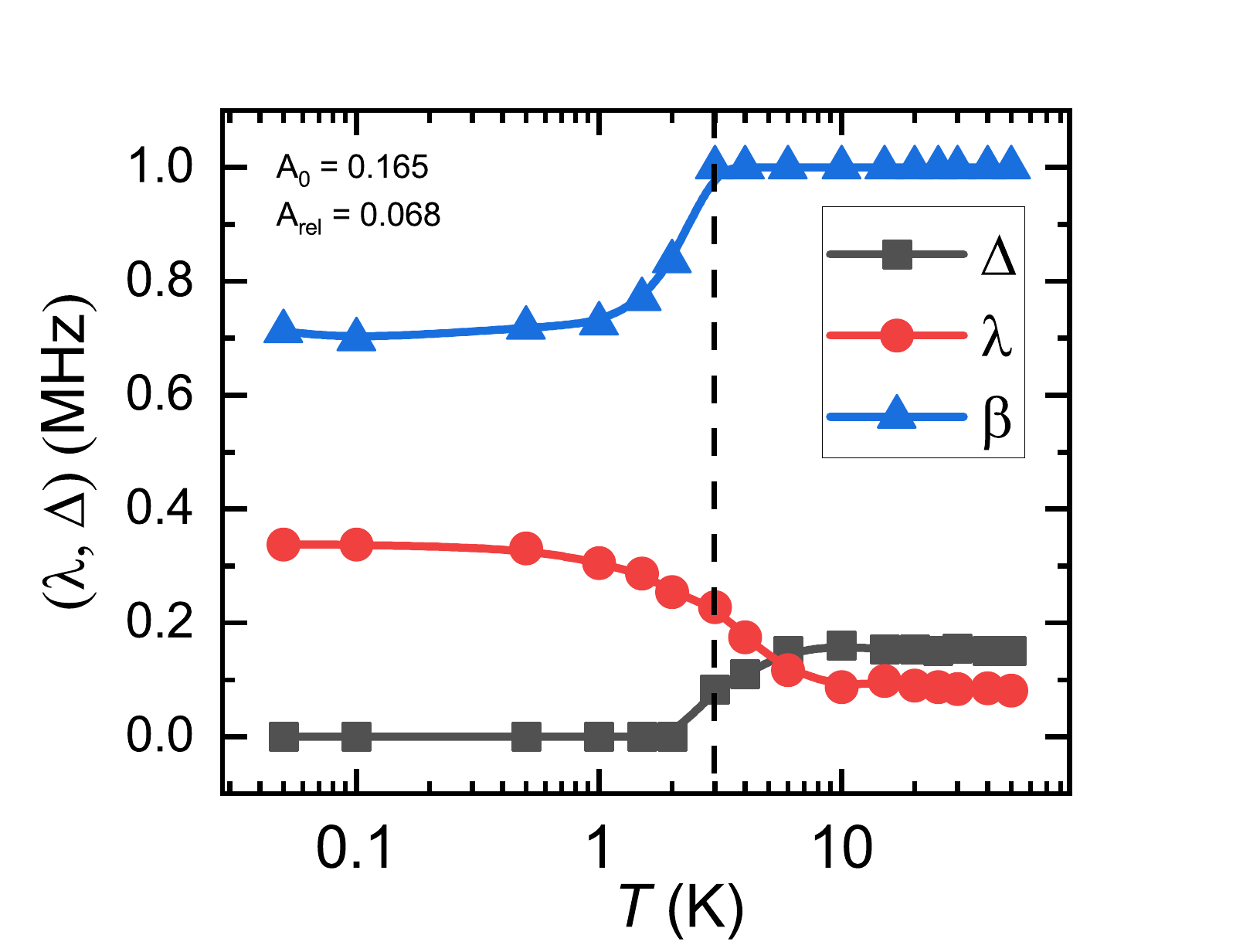}
	\caption{The temperature variation of muon spin relaxation rate ($\lambda$), static nuclear field distribution ($\Delta$) and stretching exponent ($\beta$) are shown.}
	\label{Delta_Lamda_CLRO}
\end{figure}

\begin{figure}
	\centering
	\includegraphics[width=1.0\linewidth]{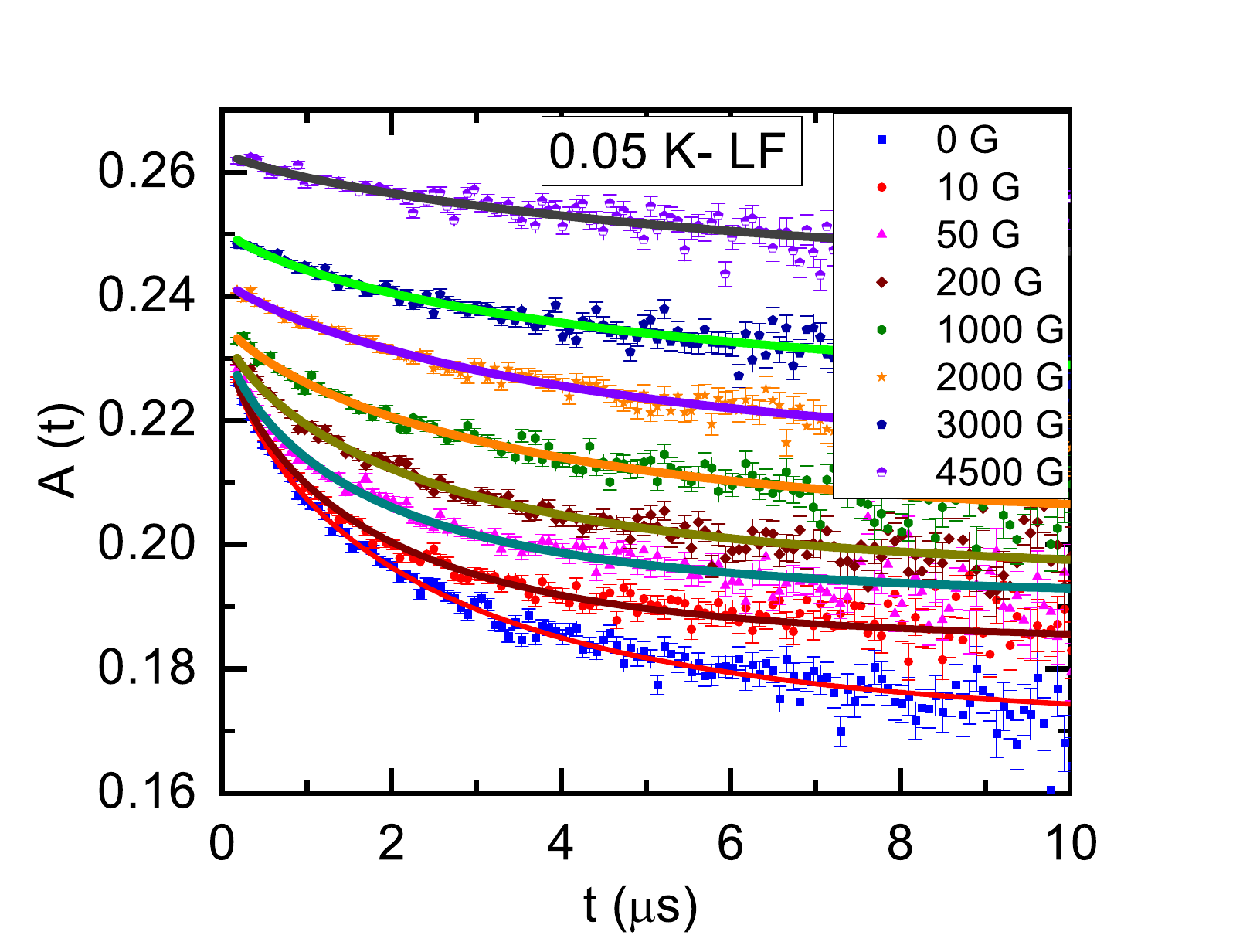}
	\caption{The variation of muon asymmetry with time at various longitudinal field (0 to 4500 G). Solid line represents fitting as described in text.}
	\label{LF_asym_CLRO}
\end{figure}

\begin{figure}[h]
	\centering{}\includegraphics[scale=0.33]{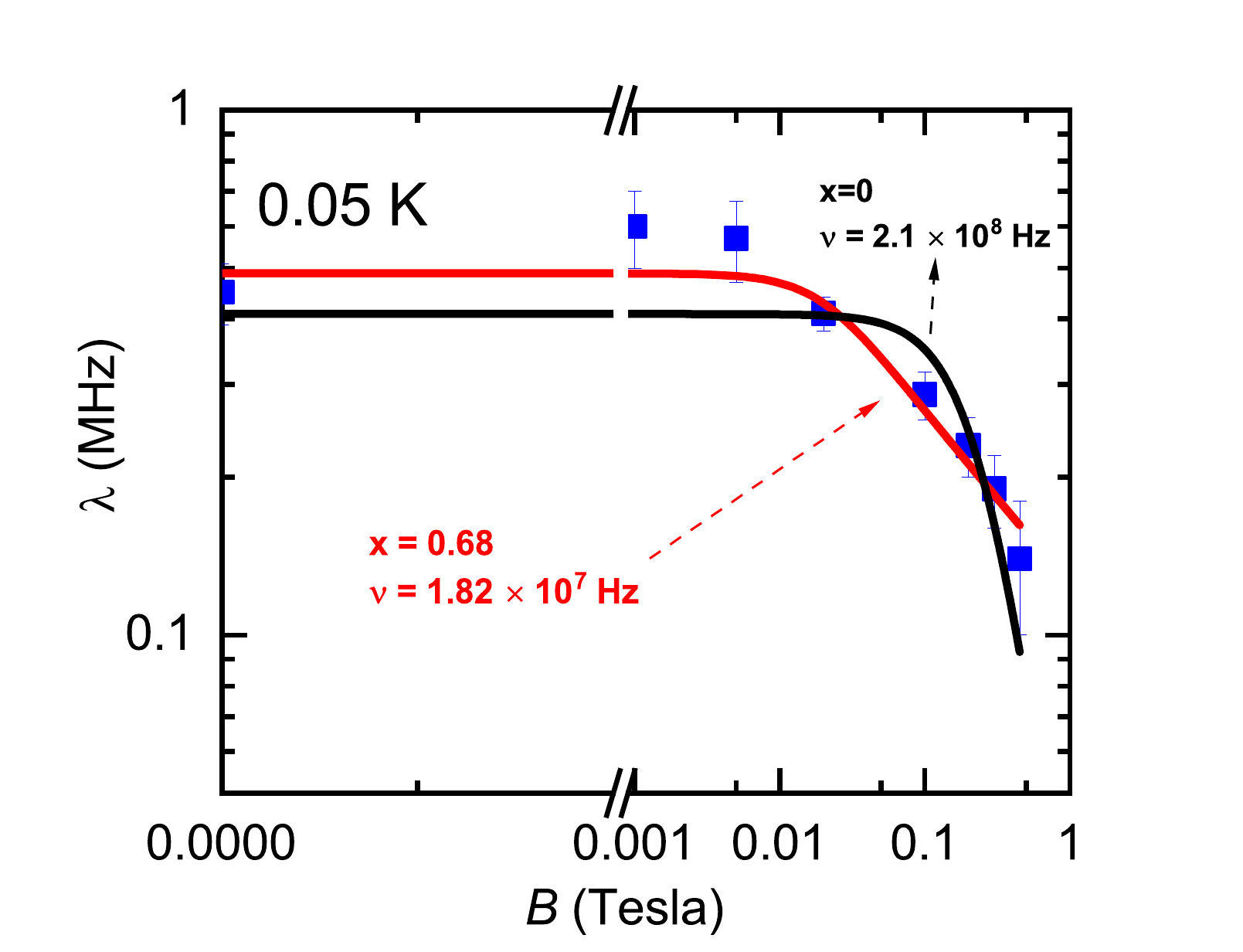}\caption{\label{fig:lamda_H_CLRO}{\small{}The variation of the muon relaxation rate at 50 mK is shown as a function of the longitudinal magnetic field for \ch{Cu3LiRu2O6}. The red curve is a fit to Equation \ref{lamda_H} with $x$ as a fitting parameter and the black curve is a fit to the same equation but with $x = 0$. }}
	\label{lamda_H_CLRO}
\end{figure}

From the data in figure \ref{LF_asym_CLRO}, it can be seen that even in a large field of 4500 G, the muons are still not decoupled from the internal fields. This indicates that the moments remain dynamic down to the lowest temperatures studied (50 mK). These are typical signatures seen in quantum spin liquid materials. The LF data were also fit to the product of the KT function with an exponential in addition to a constant background. The muon depolarization rate thus obtained is plotted as a function of the magnetic field as shown in Figure \ref{lamda_H_CLRO}.

At very high fields one expects muons to totally decouple
from the internal fields resulting in no relaxation.
But as seen in Figure \ref{LF_asym_CLRO}, a field of 4500 G is not at all sufficient to make $\lambda$ near negligible. Following the analysis of the field dependence of $\lambda$ as in Ref. \cite{YMGO}, we fit the
data to the following equation:

\begin{equation}
	\lambda (H) = 2 \Delta ^2 \tau ^x \int_{0}^{\infty} t^{-x} exp(- \nu t) Cos (\gamma_{\mu} H t) dt
	\label{lamda_H}
\end{equation}

where $\nu$ is the fluctuation frequency of local moments and
$\Delta$ is the distribution width of the local magnetic fields.
The muon gyromagnetic ratio is 
$\gamma_{\mu}$ = 2$\pi$ x 135.5342
MHz/T. A fit with $x = 0$ (black curve in Figure \ref{lamda_H_CLRO})
which implies an exponential auto-correlation function
$S(t) \sim exp(-\nu t)$ does not fit the data well and rather
$S(t) \sim (\tau/t)^x exp(- \nu t)$ is needed to fit the data. The
red solid curve is a fit to Equation \ref{lamda_H} and gives $x = 0.68$ and $\nu = 1.82 \times 10^7$ Hz. $\tau$ is the early time cut-off and is fixed
to $10^{-12} s$. This result implies the presence of long-time spin correlations but without any static order; a hallmark of spin liquids. The LF dependence shows a striking resemblance to that observed in \ch{YbMgGaO4} \cite{YMGO} and \ch{Sr3CuSb2O9} \cite{SKundu2020}, characterized by a local moment fluctuation frequency of approximately 18 MHz and the existence of spin correlations over long-time. The qualitative and quantitative results obtained through $\mu$SR analysis of CLRO align with those observed in other quantum spin liquid (QSL) systems\cite{SKundu2020}.

\bibliographystyle{apsrev4-2}
\bibliography{citation}